# Electronic Noise of a Single Skyrmion

Kang Wang,* Yiou Zhang, Vineetha Bheemarasetty, See-Chen Ying, and Gang Xiao†

*Department of Physics, Brown University, Providence, Rhode Island 02912, USA*

## Abstract

To enable the practical use of skyrmion-based devices, it is essential to achieve a balance between energy efficiency and thermal stability, while also ensuring reliable electrical detection against noise. Understanding how a skyrmion interacts with material disorder and external perturbations is thus essential. Here we investigate the electronic noise of a single skyrmion under the influence of thermal fluctuations and spin currents in a magnetic thin film. We detect the thermally induced noise with a $1/f^\gamma$ signature in the strong pinning regime but a random telegraph noise in the intermediate pinning regime. Both the thermally dominated and current-induced telegraph-like signals are detected in the weak pinning regime. Our results provide a comprehensive electronic noise picture of a single skyrmion, demonstrating the potential of noise fluctuation as a valuable tool for characterizing the pinning condition of a skyrmion. These insights could also aid in the development of low-noise and reliable skyrmion-based devices.

* kang_wang@brown.edu
† gang_xiao@brown.edu

I.    INTRODUCTION

Noise fluctuation is conveniently analyzed through its power spectrum

$$S(\omega) = \left| \int_{t_1}^{t_2} x(t) e^{-i2\pi\omega t} dt \right|^2, \qquad (1)$$

where $x(t)$ is the time-domain signal. The power spectrum possesses various signatures which may provide information about the nature of the dynamics. For example, a broadband $1/f^\gamma$ ($f = 2\pi\omega$) noise may be detected. While the noise is white when $\gamma = 0$, the $1/f^2$ signature ($\gamma = 2$) corresponds to Brownian noise and can be produced by the trajectories of a Brownian walk. According to van der Ziel's picture [1-3], the $1/f^\gamma$ signature may arise from a collection of non-identical random telegraph noise (RTN) oscillators with the form

$$S_x(f) = \frac{\Delta x^2}{2f_0} \frac{1}{1 + (\pi f/f_0)^2} \qquad (2)$$

where $f_0$ is the average fluctuation rate and $\Delta x$ is the difference of the quantity $x$ between the two states. In addition to broadband noise, there may also be a narrow band noise in the detection which produces peaks at characteristic frequencies that correspond to the length scales of the system [4-8].

Noise fluctuation has proven to be a useful method for characterizing condensed matter states such as superconducting vortices [4,5,9-12], quantum fluctuations [13-18], and charge density waves [19-24]. It could be equally useful for characterizing skyrmions. Magnetic skyrmions with a fixed chirality have been observed in non-centrosymmetric bulk magnets [25-27] and magnetic multilayers [28-32] where the antisymmetric Dzyaloshinskii-Moriya interaction [33,34] (DMI) exists and favors one sense of rotation of the magnetization over the other. Magnetic skyrmions can serve as effective agents for next-generation beyond-CMOS data storage [35-38], logic [39-41], probabilistic computing [42-45], and neuromorphic computing [46] devices owing to their non-volatility, stability, and efficient controllability.

For device applications, it is necessary to manipulate a single skyrmion via spin currents [35,36,39,40,45,46], magnetic fields, or microscopic thermal fluctuations [42-45]. This must be achieved in an efficient manner while recognizing that there is a trade-off between energy efficiency and thermal stability. Furthermore, one must be able to transduce a single skyrmion into detectable electrical signals against noise. These requirements demand an understanding how a skyrmion interacts with material disorder and external perturbations. Alternatively, one may take advantage of the noise in devices such as a skyrmion true random number generator [45]. A comprehensive study of the noise properties of a skyrmion is therefore crucial and may provide fundamental guidance on how to fabricate devices that demand low-noise performance and reliable electrical detection as well as those that utilize the noise fluctuation of a skyrmion.

The noise characterization of skyrmions has been limited to particle-based models [6,47] and only recently has been studied experimentally for the motion of a skyrmion lattice in a bulk magnet with a B20 crystal structure [7,8]. It was found that the collective transport of a skyrmion lattice produces a narrow band noise with the washboard frequency corresponding to the time required for a skyrmion particle to move one period. The electronic noise of a single skyrmion, however, remains elusive. The formation of a skyrmion configuration is a result of the delicate balance between competing energies. In materials with disorder, the size and shape of a skyrmion can vary spatially [48]. The shape of the encircling domain wall around the skyrmion core is governed by the competition between "skyrmion surface" tension energy [49,50] and local pinning energy [48,51]. Additionally, external perturbations can induce spin torques on the magnetization, causing local fluctuations that generate noise in the skyrmion.

In this work, we perform a systematic study of the electronic noise produced by a single skyrmion in response to thermal fluctuations and spin currents over a wide range of pinning conditions. We detect and analyze distinct noise signatures from the thermally dominated and current-induced dynamics of a skyrmion. This study helps us understand how a skyrmion interacts with material disorder and perturbations, which is also key to understanding the noise dynamics of other structures in physics that share similar behaviors.

## II. MAGNETIC THIN FILMS AND MAGNETIC STRUCTURES

For this study, we fabricate perpendicularly magnetized multilayers of substrate/Ta/Co$_{40}$Fe$_{40}$B$_{20}$(0.95 nm)/MgO(1.6 nm)/TaO$_x$ on thermally oxidized silicon wafers [Supplementary Note I]. The magnetic configuration in this system is determined by the competition between multiple energy terms including the perpendicular magnetic anisotropy (PMA) [52,53], exchange interaction, interfacial DMI [33,34] and Zeeman energy, and can be tuned by an applied perpendicular magnetic field $H_z$ [Supplementary Note III and Supplementary Fig. 3]. The interfacial DMI leads to chiral magnetic domain walls and stabilizes skyrmions in this system at room temperature and in a suitable field range [Supplementary Notes III and IV].

We implement moderate pinning strengths by regulating the DC power ($P_{Ta}$) for deposition of the Ta layer [Supplementary Note V]. We identify the three growth scenarios with $P_{Ta} = 3, 4,$ and 5 Watt as corresponding to the strong, intermediate, and weakly pinned samples, respectively [Supplementary Note V and Supplementary Movies 1 – 3]. Distinct pinning conditions may result from the effects of $P_{Ta}$ on thin-film structures and the variation in material parameters [Supplementary Note V]. The competition between the pinning energy, thermal energy and the current determines the noise dynamics of a skyrmion, as shown in Fig. 1(a) and discussed in more detail below.

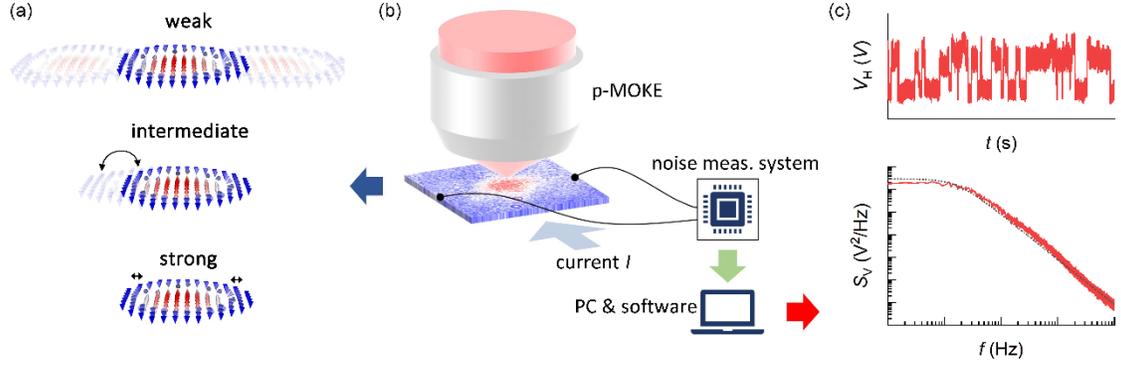

FIG. 1. Scheme of the experimental setup for studying noise of a skyrmion. (a) A schematic depiction of pinning effects on the noise dynamics of a skyrmion. In the strong pinning regime, the skyrmion is strongly pinned while multiple internal domain-wall hopping oscillations (double-headed arrows) may exist. In the intermediate pinning regime, one case is presented where one part of a skyrmion is more strongly pinned while the other part, under thermal effects, fluctuates in time between two weaker pinning sites leading to a random telegraph noise signature. In the weak pinning regime, the skyrmions hop between pinning sites more easily and both temperature and current can strongly affect the skyrmion dynamics. (b) Schematic of the experimental setup. Electronic noise of a skyrmion is studied by an electronic noise measurement system with the help of direct imaging using a polar magneto-optic Kerr effect (p-MOKE) microscope. (c) An example of the Hall voltage $V_H(t)$ and its power spectrum $S_V(f)$ obtained by the noise measurement system.

To study the electronic noise of a skyrmion, we nucleate a single skyrmion in a $20 \times 20$ μm² Hall cross and measure the magnetically induced Hall-signal fluctuations [Supplementary Note II and Fig. 1(b)]. In this case, the $x$ in Eqs. (1) and (2) is defined as the Hall voltage, $V_H = R_H I$, where $R_H$ is the Hall resistance. Notably, $R_H$ is dominated by the anomalous Hall resistance that is proportional to the perpendicular magnetization, while the conventional Hall resistance proportional to $H_z$ can be neglected in comparison [54]. We measure the noise under an applied current $I$ and a perpendicular magnetic field $H_z$. The electric current is converted into a spin current through the spin Hall solid, Ta, which exerts spin-orbit torques on the skyrmion [55-57]. This study provides us with both amplitude and frequency details of the noise of a skyrmion in responses to the external perturbations.

## III. ELECTRONIC NOISE OF A SKYRMION IN DIFFERENT PINNING REGIMES

We summarize in Fig. 2 the electronic noise results of a single skyrmion in the strong, intermediate, and weak pinning regimes. All results are measured in the temperature range from 298.9 to 319.5 K and in the current range from 0.01 to 1.0 mA. As described in more detail below, we detect a thermally induced noise with a $1/f^\gamma$ signature in the strong pinning regime [Fig. 3] but a RTN in the intermediate pinning regime [Fig. 4]. Both the thermally dominated and current-induced telegraph-like signals are detected in the weak pinning regime [Fig. 5].

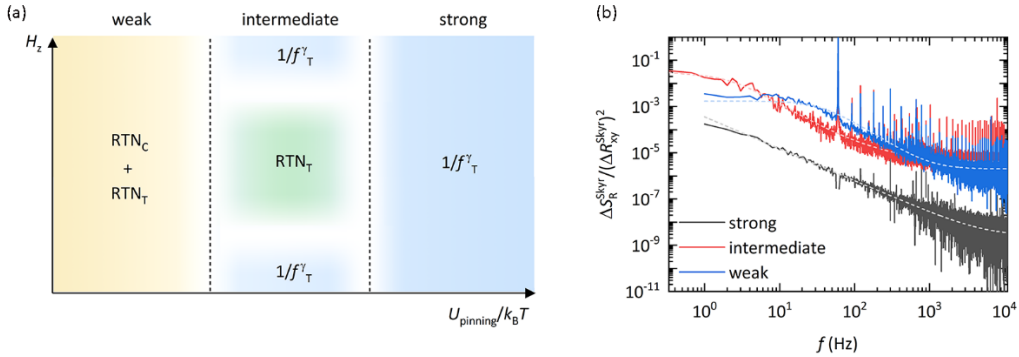

FIG. 2. (a) Summary of electronic noise of a skyrmion in different pinning regimes. The subscripts "T" and "C" represent the noise of thermally dominated or current-induced dynamics, respectively. We detect a thermally induced noise with a $1/f^\gamma$ signature in the strong pinning regime but a RTN in the intermediate pinning regime. Both the thermally induced and current-induced telegraph-like signals are detected in the weak pinning regime. (b) Normalized Hall-resistance noise $\Delta S_R^{Skyr}/(\Delta R_{xy}^{Skyr})^2$ of a skyrmion in different pinning regimes. $\Delta R_{xy}^{Skyr}$ and $\Delta S_R^{Skyr}$ represent average contributions of each skyrmion to the anomalous Hall resistance and Hall-resistance noise, respectively. In the intermediate and weak pinning regimes, we make the approximation that $\Delta S_R^{Skyr} \approx S_R$ for $N_{Skyr} = 1$. This is justified as the $S_R$ for $N_{Skyr} = 1$ significantly exceeds the background noise level, which can be disregarded in our calculations.

Figure 2(b) summarizes the amplitude details of the noise of a skyrmion for each pinning region. Instead of presenting noise $S_R$, we present a normalized Hall-resistance noise $\Delta S_R^{Skyr}/(\Delta R_{xy}^{Skyr})^2$ where $\Delta R_{xy}^{Skyr}$ and $\Delta S_R^{Skyr}$ are average contributions of each skyrmion to the anomalous Hall resistance and Hall-resistance noise, respectively, as discussed below. The normalized noise accounts for the variations in the Hall resistance of a single skyrmion across different samples and therefore allows us to compare different samples.

A. $1/f^\gamma$ noise of a skyrmion in the strong pinning regime

In the strong pinning regime, skyrmions move negligibly owing to the lower thermal energy relative to the pinning energy $U_{pinning}$. Still, the internal domain-wall hopping exists and may contribute to the $1/f^\gamma$ noise signature [Fig. 3(b), also see Supplementary Note VI and Supplementary Figs. 7 and 8]. The speculation of internal domain-wall hopping oscillations contributing to the $1/f^\gamma$ noise of a skyrmion finds validation through micromagnetic simulations as depicted in Supplementary Note VII and Supplementary Fig. 9. The pinning energy $U_{pinning}$ governs the domain-wall hopping through the Arrhenius law $f_0 = f_{00}\exp(-U_{pinning}/k_B T)$, where $f_{00}$ is the attempt frequency. The $U_{pinning}$ distribution therefore affects the distribution of the internal domain-wall hopping oscillators as well as the noise signature from this collection of non-identical oscillators. Conversely, the distribution of $U_{pinning}$ can be inferred through an examination of the noise spectrum. When the distribution is uniform, the value of $\gamma$ equals 1. A fitting of the $1/f^\gamma$ noise with $\gamma = 1.36$ [gray curve in Fig. 3(b)] uncovers the $U_{pinning}$ distribution contributing to the $1/f^\gamma$ noise, as depicted in Fig. 3(a). For further fitting specifics, refer to Supplementary Note VIII.

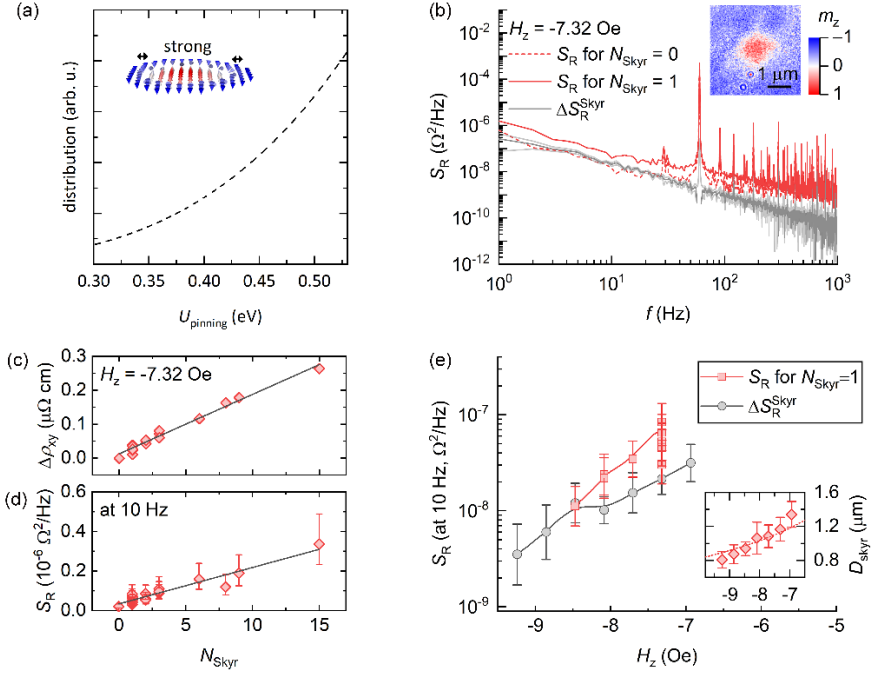

FIG. 3. $1/f^\gamma$ noise of a skyrmion in the strong pinning regime. (a) A plot illustrating distribution of the $U_{\text{pinning}}$ that contribute to the $1/f^\gamma$ noise of a skyrmion. The result is obtained by fitting the gray curve in (b), which exhibits a $1/f^\gamma$ signature with $\gamma = 1.36$. The inset provides a schematic depiction of the noise dynamics of a skyrmion under strong pinning, where the skyrmion is strongly immobilized while potentially undergoing multiple internal domain-wall hopping oscillations (double-headed arrows). (b) Red dashed line: $S_R$ spectra with the $1/f^\gamma$ noise signature for $N_{\text{Skyr}} = 0$ at $H_z = -7.32$ Oe. Red solid line: $S_R$ spectra with the $1/f^\gamma$ noise signature for $N_{\text{Skyr}} = 1$ at $H_z = -7.32$ Oe. Gray line: an average contribution of each skyrmion to the noise $\Delta S_R^{\text{Skyr}}$ where the background noise is subtracted. The inset shows a p-MOKE image of the skyrmion for noise measurements. (c) The relative Hall resistivity $\Delta \rho_{xy}$ as a function of $N_{\text{Skyr}}$. (d) $S_R$ at 10 Hz as a function of $N_{\text{Skyr}}$. The dotted line is a linear fit to $S_R = S_{R,0} + \Delta S_R^{\text{Skyr}} N_{\text{Skyr}}$. (e) $S_R$ for $N_{\text{Skyr}} = 1$, $\Delta S_R^{\text{Skyr}}$ and the skyrmion diameter $D_{\text{Skyr}}$ as a function of $H_z$. Dotted line in the inset represents the fit to the experimental data using a theoretical model [58] [Supplementary Note II].

To eliminate the background noise, we study the noise of discrete skyrmions at a constant field [Supplementary Note VI and Supplementary Fig. 7]. At a field where skyrmions are stabilized, multiple states with variable skyrmion numbers $N_{\text{Skyr}}$ may be accessed through a field cycling process [Supplementary Note IV]. Figure 3(c) shows a linear variation of the relative Hall resistivity $\Delta\rho_{xy}$ with $N_{Skyr}$. Each skyrmion contributes to the $\Delta\rho_{xy}$ in the amount of $\Delta\rho_{xy}^{Skyr} = 17.6 \pm 1.25$ nΩ cm. This aligns closely with the previously reported electrical detection of an isolated skyrmion using the anomalous Hall effect [59]. In addition, the Hall-resistance noise $S_R$ also increases linearly with $N_{Skyr}$ [Fig. 3(d)]. Fitting to $S_R = S_{R,0} + \Delta S_R^{Skyr} N_{Skyr}$ yields the average contribution of each skyrmion to the noise $\Delta S_R^{Skyr}$ as well as the contribution from the uniform magnetization state $S_{R,0}$. This fitting assumes negligible interactions between skyrmions due to their large spatial separation [Supplementary Note VI]. Although $\Delta S_R^{Skyr}$ is smaller than the background noise by one order of magnitude [Supplementary Fig. 7], it is distinguishable and clearly exhibits the $1/f^\gamma$ signature [gray line in Fig. 3(b)]. Analogous measurements performed at different fields for this sample and other samples all concur with the $1/f^\gamma$ signature with $1.0 < \gamma < 1.4$, independent of the skyrmion's polarization [Supplementary Note VI and Supplementary Fig. 8].

B.   Field-dependent $1/f^\gamma$ noise of a skyrmion

Figure 3(e) displays the field-dependent noise of a skyrmion at 10 Hz as well as the skyrmion diameter $D_{Skyr}$. It is noted that while the skyrmion size decreases by about 43% from −6.93 to −9.24 Oe [inset in Fig. 3(e)], the $\Delta S_R^{Skyr}$ decreases by almost two orders of magnitude. This is contrary to what one would expect, which is that the number of internal domain-wall hopping oscillators is proportional to the total length of domain walls which would thereby linearly affect the noise amplitude [Supplementary Note IX and Supplementary Fig. 10]. The significant reduction in noise with respect to the skyrmion size suggests that the number of internal domain-wall hopping oscillators in a skyrmion decreases significantly with a reduction in skyrmion size. It can be inferred that smaller skyrmions are more appropriate for applications due to their substantially lower noise levels.

We calculate the signal-to-noise ratio (SNR) by using $\sqrt{\Delta S_R^{Skyr}}/\Delta R_{xy}^{Skyr}$ where $\Delta R_{xy}^{Skyr}$ is the average contribution of each skyrmion to the anomalous Hall resistance. The SNR of a skyrmion is on the order of 2.0% Hz$^{-0.5}$ at 1 Hz and 0.5% Hz$^{-0.5}$ at 10 Hz, as evidenced by SNR$^2 = \Delta S_R^{Skyr}/\left(\Delta R_{xy}^{Skyr}\right)^2$ in Fig. 2(b). A low SNR is crucial in the electrical detection of a skyrmion owing to the small signal from this tiny magnetic structure [46,59].

C.   Random telegraph noise of a skyrmion in the intermediate pinning regime

In the intermediate pinning regime, a skyrmion exhibits the ability to hop between certain pinning sites with greater ease yet remains strongly pinned by specific pinning centers. [Supplementary Note X, Supplementary Fig. 11, and Supplementary Movie 4]. The case presented here involves one part of a skyrmion being more strongly pinned while the other part, under thermal effects, fluctuates in time between two weaker pinning sites [Figs. 4(a) and (b)]. This behavior is schematically represented by the bended double-headed arrow in Fig. 4(a). This produces a RTN signature [Fig. 4(c), also see Ref. [45]]. Along with the RTN, we identify a fluctuation in $\Delta\rho_{xy}$ over time between two discrete states and correspondingly a skyrmion-size variation between a small-skyrmion ($\Delta\rho_{xy} \approx 0$) and a large-skyrmion ($\Delta\rho_{xy} \approx 8$ n$\Omega$ cm) configuration [Fig. 4(d) and Supplementary Movie 5]. This shares the same physics as the internal domain-wall hopping oscillations in the strong pinning regime. There may also be more internal domain-wall hopping oscillations in the skyrmion, depicted by the straight double-headed arrow in Fig. 4(a). Nevertheless, these oscillations manifest with significantly smaller amplitudes $\Delta V_H^2/2I^2$. Their contributions to the overall noise [$10^{-7} - 10^{-6}$ $\Omega^2$/Hz at 1 Hz as shown by the gray curve in Fig. 4(c)] are negligible in comparison to the detected noise in the intermediate pinning regime [$10^{-5}$ $\Omega^2$/Hz at 1 Hz as shown by the red curve in Fig. 4(c)].

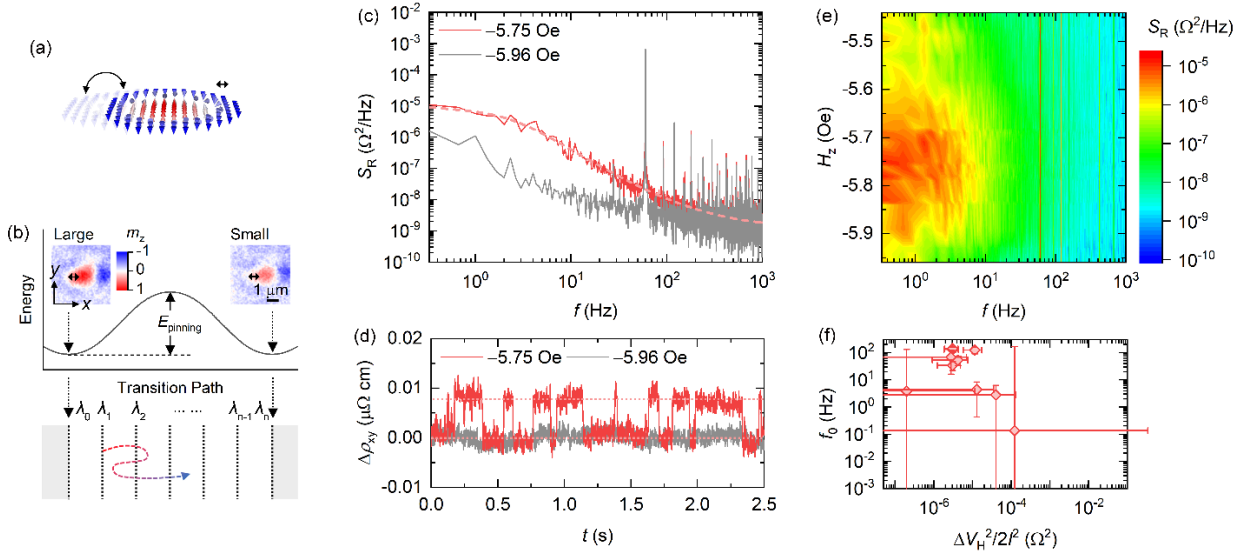

FIG. 4. Random telegraph noise of a skyrmion in the intermediate pinning regime. (a) Schematic depiction of skyrmion dynamics in the intermediate pinning regime. The case presented here involves one part of a skyrmion being more strongly pinned while the other part, under thermal effects, fluctuates in time between two weaker pinning sites leading to a random telegraph noise signature. (b) Energy path for the two-state transition. The small-skyrmion and large-skyrmion states occur at local minima in the energy landscape separated by an energy barrier. The transition between the two states involves passing through a series of interfaces, $\{\lambda_L = \lambda_0, \lambda_1, \cdots, \lambda_{n-1}, \lambda_n = \lambda_S\}$, defined as isosurfaces of a monotonically varying order parameter, $\zeta$, such as the perpendicular magnetization $m_z$. Due to the presence of thermal energy, recrossings between neighboring interfaces are inevitable, as illustrated by the dashed arrow in the schematic. (c) Hall-resistance noise spectra $S_R$ of a skyrmion in the intermediate pinning regime. The spectra are measured at $H_z = -5.75$ Oe (red) and $-5.96$ Oe (gray) where the skyrmion fluctuates in time between the two states or remains in a deterministic small-skyrmion state, respectively. The dashed line is a fit to the RTN plus a $1/f^\gamma$ noise. (d) Relative Hall-resistivity $\Delta\rho_{xy}$ variations in time of a skyrmion at $H_z = -5.75$ Oe (red) and $-5.96$ Oe (gray). (e) Map of Hall-resistance noise spectra $S_R$ at different fields. A transition of the RTN into the $1/f^\gamma$ noise is observed by increasing or decreasing the field to the deterministic small-skyrmion and large-skyrmion state, respectively. (f) The fluctuation rate $f_0$ as a function of the amplitude $\Delta V^2/2I^2$ of the RTN of isolated skyrmions in the intermediate pinning regime. All results are measured at 307.1 K and with a current $I = -0.2$ mA. This current corresponds to the current density of $-2.4 \times 10^9$ A/m² flowing in the Ta buffer layer. The red curve in (d) is reproduced from Ref. [45].

The two identified states, referred to as the "small-skyrmion" and "large-skyrmion" states, correspond to local minima in the energy landscape, separated by an energy barrier, as illustrated in Fig. 4(b). The fluctuation between these states is thermally induced, and the fluctuation rate is determined by the competition between the pinning energy barrier $U_{\text{pinning}}$ and the thermal energy $k_B T$. Our previous work has shown that the energy landscape can be influenced by adjusting the applied magnetic field and a spin current through the Zeeman energy and spin-orbit torques, respectively [45]. This enables the manipulation of the local dynamics of a skyrmion and, in turn, influences the noise generated by the system [45]. The skyrmion is more likely being in the small-skyrmion state when the applied field $H_z$ is increased or when the current is applied along the +x axis in Fig. 4(b) [45]. Similarly, the large-skyrmion state is more probable with decreasing the applied field or applying a current in the −x axis [45]. Along with the deterministic small-skyrmion and large-skyrmion configuration at a higher or lower field, respectively, a $1/f^\gamma$ noise signature is detected [Figs. 4(c) and (e)]. This closely resembles the behavior of a strongly pinned skyrmion.

D.  Effects of the energy landscape on the internal domain-wall hopping oscillation dynamics

We observe and analyze the RTN for multiple isolated skyrmions where the fluctuation rate $f_0$ varies between skyrmions [Fig. 4(f), also see Supplementary Note X and Supplementary Fig. 12]. This fluctuation rate variation is a result of different energy paths for the two-state transition and provides us with a platform to study how the energy landscape affects the internal domain-wall hopping oscillation dynamics of a skyrmion. The Arrhenius law $f_0 = f_{00}\exp(-U_{\text{pinning}}/k_B T)$ tells us how $U_{\text{pinning}}$ and the thermal energy $k_B T$ affect the RTN dynamics. Additionally, the attempt frequency $f_{00}$ may carry an activation entropy which implies that a longer path must be explored more randomly [60-62]. It has been identified recently through simulation studies that the entropic effect is crucial in describing other processes including the magnetization switching [60,62] and skyrmion annihilation [61]. The entropic effect can be elucidated through the forward flux sampling method, which involves a series of interfaces $\{\lambda_0 = \lambda_L, \lambda_1, \cdots, \lambda_{n-1}, \lambda_n = \lambda_S\}$ in configuration space between the large-skyrmion and small-skyrmion states. These interfaces are defined as isosurfaces of a monotonically varying order parameter, $\zeta$, such as the perpendicular magnetization $m_z$ [Fig. 4(b)]. Despite the energy profile being characterized by two local minima and an energy barrier in between, recrossings between neighboring interfaces are inevitable owing to the thermal energy. The fluctuation rate is then expressed as $\prod_{i=0}^{n-1} P(\lambda_{i+1}|\lambda_i)$, where $P(\lambda_{i+1}|\lambda_i)$ represents the probability that a trajectory originating from $\lambda_i$ reaching $\lambda_{i+1}$ before returning to the basin initial state. For a large skyrmion-size variation, more interfaces are required to transition from the basin large-skyrmion to the small-skyrmion state, reducing $\prod_{i=0}^{n-1} P(\lambda_{i+1}|\lambda_i)$ and, consequently, the fluctuation rate. We illustrate in Fig. 4(e) that $f_0$ increases when decreasing $\Delta V_H^2/2I^2$, with $\Delta V_H^2/2I^2$ reflecting the skyrmion-size variation. This provides evidence that, in addition to the $U_{\text{pinning}}$ and thermal energy $k_B T$, the entropic effect is also crucial in the noise dynamics of a skyrmion.

We note that the RTN transitions into the $1/f^\gamma$ noise by increasing $N_{\text{Skyr}}$ [Supplementary Note X and Supplementary Fig. 13]. Each skyrmion represents a RTN oscillator. A collection of multiple skyrmions represents a system that verifies the $1/f^\gamma$ signature arising from a collection of multiple internal domain-wall hopping oscillators.

E.     Thermally induced and current-induced telegraph-like noise in the weak pinning regime

In the weak pinning regime, skyrmions exhibit enhanced mobility as they readily hop between pinning sites [Fig. 5(a), also see Supplementary Movie 2 and Supplementary Fig. 3]. In this scenario, we are unable to detect the electronic noise of a specified skyrmion. In addition to the thermal fluctuation, the current may also play a significant role in the dynamics.

In this section, we begin with displaying the $S_R$ measured at $H_z = -3.58$ Oe where multiple skyrmions are stabilized [Supplementary Movie 6] to illustrate the noise signature and underlying physics [Figs. 5(b) – (e)]. It is observed that skyrmions fit commensurably into the Hall cross at this field.

However, this commensurability effect is not observed in the strong and intermediate pinning regimes, where the skyrmion order is constrained by the pinning sites.

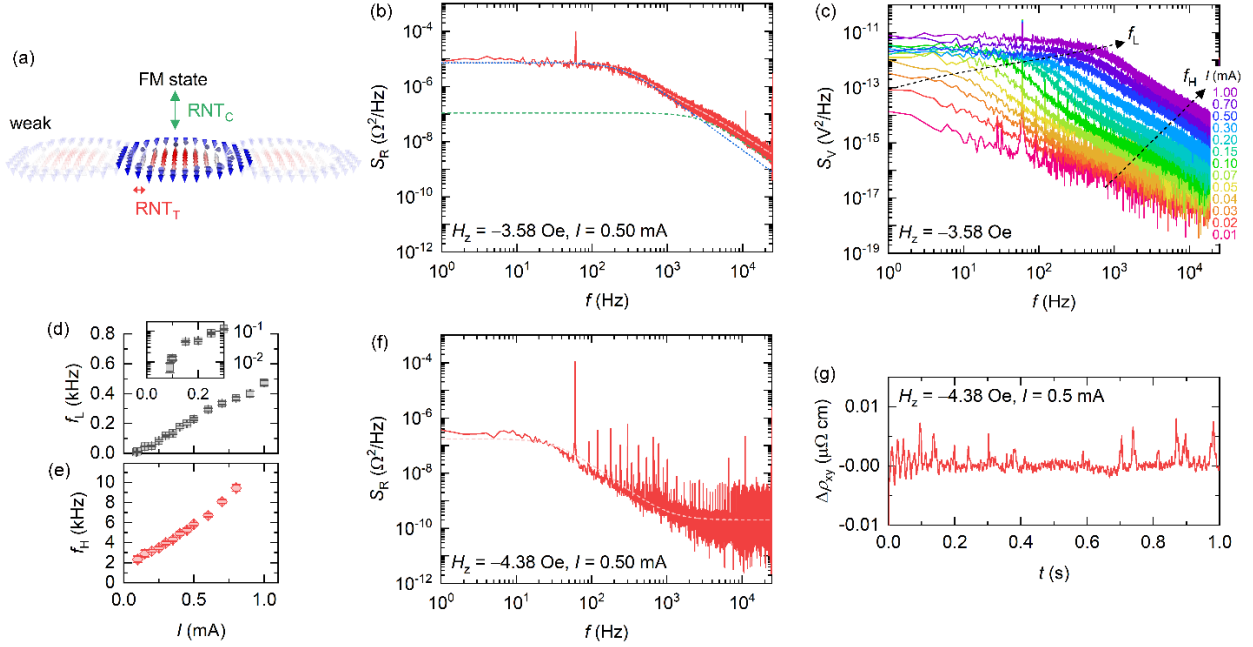

FIG. 5. Thermally induced and current-induced telegraph-like noise of a skyrmion in the weak pinning regime. (a) Schematic depiction of the skyrmion dynamics in the weak pinning regime. Skyrmions exhibit enhanced mobility as they readily hop between pinning sites. Both temperature and current can strongly affect the skyrmion dynamics. (b) Hall-resistance noise $S_R$ measured at $H_z = -3.58$ Oe where skyrmions are stabilized. The dash-dotted line is a fit to Eq. (3). The dotted blue and dashed green lines are fits to the lower-frequency and higher-frequency telegraph-like noise, respectively. (c) Hall-voltage noise $S_V$ measured at various currents. (d), (e) The central frequency $f_L$ (d) and $f_H$ (e) of the lower-frequency and higher-frequency telegraph-like noise as a function of current $I$. The inset in (d) presents a semi-logarithmic plot to highlight the low current range. 1 mA current corresponds to the current density of $1.2 \times 10^{10}$ A/m$^2$ flowing in the Ta buffer layer. (f) Hall-resistance noise $S_R$ measured at $H_z = -4.38$ Oe. (g) relative Hall-resistivity $\Delta\rho_{xy}$ variations in time at $H_z = -4.38$ Oe where only a single skyrmion may be nucleated. Nucleation of a skyrmion leads to an increase in $\Delta\rho_{xy}$ which reduces back to zero when the skyrmion propagates out of the Hall cross. All results are measured at 319.5 K.

We observe two telegraph-like noise signals in Fig. 5(b) which can be well fitted to,

$$S_V = \frac{S_{V,0}^{1Hz}}{f^\gamma} + \frac{\Delta V_{H,L}^2/2f_L}{1+(\pi f/f_L)^2} + \frac{\Delta V_{H,H}^2/2f_H}{1+(\pi f/f_H)^2}, \tag{3}$$

where $\Delta V_{H,L}^2/2I^2$ and $\Delta V_{H,H}^2/2I^2$ are amplitudes of the lower-frequency and the higher-frequency telegraph-like signals, respectively. In our classification, we refer to this noise as "telegraph-like noise" instead of "RTN" because it arises from a collection of multiple internal domain-wall hopping RTN oscillators, though with a narrow distribution of these oscillators, as elaborated upon below. Notably, we confirm that the lower-frequency signal originates from the current effect as the central frequency $f_L$ decreases and approaches zero when the current is decreased to zero [Figs. 5(c) and (d)]. The higher-frequency signal is dominated by the thermal effect since the central frequency $f_H$ approaches a non-zero value (approximately 2 kHz) in the zero-current limit [Figs. 5(c) and (e)]. We note that the temperature induced by Joule heating, over the range of currents employed in our study, remain below 1 K [Supplementary Note XI and Supplementary Fig. 14], which is negligible in comparison to the range of temperatures under our investigation.

When the magnetic field is increased to $H_z = -4.38$ Oe, Hall-resistance measurements in Fig. 5(g) show that only discrete skyrmions are nucleated and annihilated by the current, as evidenced by distinct spikes. As a result, the lower-frequency RTN is visible, while the higher-frequency signal dominated by thermal effects is suppressed. These observations suggest that the lower-frequency RTN arises from the current-induced skyrmion nucleation and annihilation process, which is discussed in further detail below.

We investigate the origins of the two noise signals by probing Hall-voltage fluctuations over a smaller Hall cross with dimensions of $5 \times 5$ μm$^2$, where only a single magnetic bubble can be nucleated and propagate [inset in Fig. 6(a)]. We cannot ascertain whether the magnetic bubbles are topologically equivalent to skyrmions or not. To avoid any misleading, we refer to them as "magnetic bubbles" rather than "skyrmions". However, this distinction does not impact our understanding of the origins of the two noise signals. We nucleate a magnetic bubble at defect positions or at sample boundaries. At a small current ($\leq 0.06$ mA), the magnetic bubble domain is elongated and flows continuously through the Hall cross and only the higher-frequency telegraph-like noise is visible [Fig. 6(a)]. By contrast, a larger current ($\geq 0.07$ mA) induces discrete domain nucleation and propagation, causing the Hall voltage to fluctuate in time between two discrete states [inset in Fig. 6(b), also see Supplementary Movie 7]. This thereby results in the additional lower-frequency bump [Figs. 6(a) and (b)]. The power spectrum over a period from 0.15 to 0.22 s of the Hall signal, again, only shows the higher-frequency signal [gray line in Fig. 6(b)]. During this time, a domain continuously flows through the Hall cross. These results indicate that the lower-frequency telegraph noise is a result of current-induced bubble domain nucleation and propagation while the higher-frequency signal most likely arises from thermally dominated domain-wall fluctuations, as it is visible even with only a single bubble domain propagating through the Hall cross and has not been observed in the saturated ferromagnetic state. In the weak pinning regime, $U_\text{pinning}$ exhibits a much narrower distribution. This results in the telegraph-like noise signature [Figs. 5(b) and (c), and Figs. 6(a) and (b)], in contrast to $1/f^\gamma$ noise signature in the strong pinning regime, with the measured frequency $f_\text{H}$ being the averaged result over the domain-wall hopping oscillation dynamics.

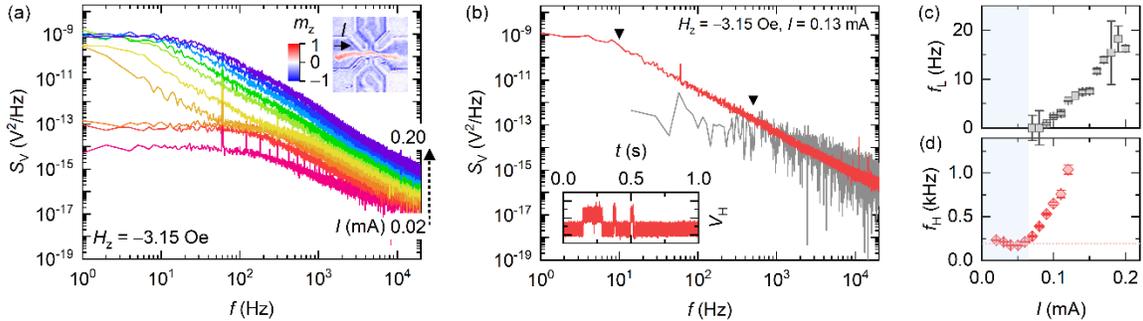

FIG. 6. Electronic noise of a bubble domain in a Hall cross with a dimension of $5 \times 5$ μm$^2$. (a) Hall-voltage noise $S_V$ measured at various currents. The inset shows a p-MOKE image of the Hall cross with a dimension of $5 \times 5$ μm$^2$ where only a single elongated magnetic bubble is nucleated and propagates through the Hall cross. (b) $S_V$ measured at $I = 0.13$ mA. The inset shows the Hall-voltage variations in time at $I = 0.13$ mA. The gray line is the power

spectral density of the period between 0.15 and 0.22 s of Hall-voltage signals. (c), (d) The central frequency $f_L$ (c) and $f_H$ (d) of the lower-frequency and higher-frequency telegraph-like noise as a function of current $I$. All results are measured at 318.3 K. In this configuration, a current of 1 mA corresponds to a current density of $4.8 \times 10^{10}$ A/m² flowing in the underlying Ta buffer layer.

## IV. DISCUSSION AND CONCLUSION

Our work has revealed comprehensive electronic noise properties of a single skyrmion. Distinctive amplitudes and frequency spectra can be mapped out according to the strong, intermediate, and weak pinning regimes. As summarized in Fig. 2, we detect a noise with a $1/f^\gamma$ signature in the strong pinning regime but a RTN in the intermediate pinning regime, and two telegraph-like signals in the weak pinning regime. Distinct noise signatures arise from the thermally dominated internal domain-wall hopping oscillation and/or the current-induced skyrmion nucleation. It is understood that the competition between the thermal and pinning energies governs the internal domain-wall hopping oscillation dynamics. Additionally, we reveal that the entropic effect is also significant in determining the dynamics and thereby the resulting noise signatures.

Our results demonstrate that noise fluctuation can serve as an insightful probe for characterizing the pinning condition of a skyrmion, with the $1/f^\gamma$ signature indicating a flatter distribution of oscillators in the strong pinning regime [Fig. 3(a)]. In contrast, a much narrower distribution leads to telegraph-like noise in the weak pinning regime.

Furthermore, our research has implications for applications that either require reliable electrical detection or utilize the inherent noise of a skyrmion. A comprehensive noise picture of a skyrmion can guide the fabrication of low-noise and reliable skyrmion-based devices. In the strong and intermediate pinning regimes, the skyrmion exhibits high thermal stability, but pinning effects impede its efficient manipulation. In contrast, a skyrmion can be efficiently manipulated in the weak pinning regime, although its thermal stability is reduced. We note that even a small current of $6.1 \times 10^9$ A/m² may cause the unwanted nucleation of a skyrmion [Figs. 5(f) and (g)], which may impose a current limit for applications. Alternatively, artificial pinning centers may be used to guide a skyrmion along a desirable path and position it at a fixed location for detection. In this way, artificial pinning centers can help to achieve the trade-off between energy efficiency and thermal stability necessary for applications. Noise fluctuation can also be an insightful probe for characterizing skyrmion interactions with artificial pinning.

Additionally, our noise study of a skyrmion may also be key to understanding the noise dynamics of other structures such as superconducting vortices [10] and charge density waves [23] that exhibit similar behaviors [Supplementary Note XII]. A complete picture connecting the noise and the dynamics of these condensed matter states remains elusive and demands further exploration.

**ACKNOWLEDGEMENTS**

G.X. acknowledges funding support from the National Science Foundation (NSF) under Grant No. OMA-1936221. Y.Z. acknowledges funding support from the Fermilab-Graduate Instrumentation Research Award from DOE Award No. DE-AC05-00OR22725. We use the Heidelberg MLA150 maskless aligner which is under the support of NSF Grant No. DMR-1827453.

**REFERENCES**

[1] A. Van Der Ziel, Physica **16**, 359 (1950).

[2] M. Weissman, Rev. Mod. Phys. **60**, 537 (1988).

[3] P. Dutta and P. Horn, Rev. Mod. Phys. **53**, 497 (1981).

[4] G. D'anna, P. Gammel, H. Safar, G. Alers, D. Bishop, J. Giapintzakis, and D. Ginsberg, Phys. Rev. Lett. **75**, 3521 (1995).

[5] Y. Togawa, R. Abiru, K. Iwaya, H. Kitano, and A. Maeda, Phys. Rev. Lett. **85**, 3716 (2000).

[6] S. A. Díaz, C. O. Reichhardt, D. P. Arovas, A. Saxena, and C. Reichhardt, Phys. Rev. B **96**, 085106 (2017).

[7] T. Sato, W. Koshibae, A. Kikkawa, T. Yokouchi, H. Oike, Y. Taguchi, N. Nagaosa, Y. Tokura, and F. Kagawa, Phys. Rev. B **100**, 094410 (2019).

[8] T. Sato, A. Kikkawa, Y. Taguchi, Y. Tokura, and F. Kagawa, Phys. Rev. B **102**, 180411 (2020).

[9] S. Okuma, J. Inoue, and N. Kokubo, Phys. Rev. B **76**, 172503 (2007).

[10] A. Marley, M. Higgins, and S. Bhattacharya, Phys. Rev. Lett. **74**, 3029 (1995).

[11] C. Olson, C. Reichhardt, and F. Nori, Phys. Rev. Lett. **81**, 3757 (1998).

[12] T. Tsuboi, T. Hanaguri, and A. Maeda, Phys. Rev. Lett. **80**, 4550 (1998).

[13] J. M. Martinis, S. Nam, J. Aumentado, and C. Urbina, Phys. Rev. Lett. **89**, 117901 (2002).

[14] E. Paladino, Y. Galperin, G. Falci, and B. Altshuler, Rev. Mod. Phys. **86**, 361 (2014).

[15] O. S. Lumbroso, L. Simine, A. Nitzan, D. Segal, and O. Tal, Nature **562**, 240 (2018).

[16] Y. Zhang and G. Xiao, Phys. Rev. B **100**, 224402 (2019).

[17] S. Larocque, E. Pinsolle, C. Lupien, and B. Reulet, Phys. Rev. Lett. **125**, 106801 (2020).

[18] O. Shein-Lumbroso, J. Liu, A. Shastry, D. Segal, and O. Tal, Phys. Rev. Lett. **128**, 237701 (2022).

[19] R. Fleming and C. Grimes, Phys. Rev. Lett. **42**, 1423 (1979).

[20] G. Grüner, A. Zawadowski, and P. Chaikin, Phys. Rev. Lett. **46**, 511 (1981).

[21] I. Bloom, A. Marley, and M. Weissman, Phys. Rev. Lett. **71**, 4385 (1993).

[22] K. Bennaceur, C. Lupien, B. Reulet, G. Gervais, L. Pfeiffer, and K. West, Phys. Rev. Lett. **120**, 136801 (2018).

[23] M. Yoshida, T. Sato, F. Kagawa, and Y. Iwasa, Phys. Rev. B **100**, 155125 (2019).

[24] A. K. Geremew, S. Rumyantsev, B. Debnath, R. K. Lake, and A. A. Balandin, Appl. Phys. Lett. **116** (2020).

[25] S. Mühlbauer, B. Binz, F. Jonietz, C. Pfleiderer, A. Rosch, A. Neubauer, R. Georgii, and P. Böni, Science **323**, 915 (2009).

[26]	X. Yu, Y. Onose, N. Kanazawa, J. H. Park, J. Han, Y. Matsui, N. Nagaosa, and Y. Tokura, Nature **465**, 901 (2010).

[27]	X. Yu, N. Kanazawa, Y. Onose, K. Kimoto, W. Zhang, S. Ishiwata, Y. Matsui, and Y. Tokura, Nat. Mater. **10**, 106 (2011).

[28]	N. Romming, C. Hanneken, M. Menzel, J. E. Bickel, B. Wolter, K. von Bergmann, A. Kubetzka, and R. Wiesendanger, Science **341**, 636 (2013).

[29]	W. Jiang *et al.*, Science **349**, 283 (2015).

[30]	C. Moreau-Luchaire *et al.*, Nat. Nanotechnol. **11**, 444 (2016).

[31]	A. Soumyanarayanan *et al.*, Nat. Mater. **16**, 898 (2017).

[32]	K. Wang, V. Bheemarasetty, J. Duan, S. Zhou, and G. Xiao, J. Magn. Magn. Mater. **563**, 169905 (2022).

[33]	I. Dzyaloshinsky, J. Phys. Chem. Solids **4**, 241 (1958).

[34]	T. Moriya, Phys. Rev. **120**, 91 (1960).

[35]	A. Fert, V. Cros, and J. Sampaio, Nat. Nanotechnol. **8**, 152 (2013).

[36]	K. Wang, L. Qian, S.-C. Ying, G. Xiao, and X. Wu, Nanoscale **11**, 6952 (2019).

[37]	J. Tang, Y. Wu, W. Wang, L. Kong, B. Lv, W. Wei, J. Zang, M. Tian, and H. Du, Nat. Nanotechnol. **16**, 1086 (2021).

[38]	W. Wang, D. Song, W. Wei, P. Nan, S. Zhang, B. Ge, M. Tian, J. Zang, and H. Du, Nat. Commun. **13**, 1593 (2022).

[39]	X. Zhang, M. Ezawa, and Y. Zhou, Sci. Rep. **5**, 9400 (2015).

[40]	Z. Yan, Y. Liu, Y. Guang, K. Yue, J. Feng, R. Lake, G. Yu, and X. Han, Phys. Rev. Appl. **15**, 064004 (2021).

[41]	K. Raab, M. A. Brems, G. Beneke, T. Dohi, J. Rothörl, F. Kammerbauer, J. H. Mentink, and M. Kläui, Nat. Commun. **13**, 6982 (2022).

[42]	J. Zázvorka *et al.*, Nat. Nanotechnol. **14**, 658 (2019).

[43]	Y. Jibiki *et al.*, Appl. Phys. Lett. **117**, 082402 (2020).

[44]	R. Ishikawa, M. Goto, H. Nomura, and Y. Suzuki, Appl. Phys. Lett. **119**, 072402 (2021).

[45]	K. Wang, Y. Zhang, V. Bheemarasetty, S. Zhou, S.-C. Ying, and G. Xiao, Nat. Commun. **13**, 722 (2022).

[46]	K. M. Song *et al.*, Nat. Electron. **3**, 148 (2020).

[47]	C. Reichhardt and C. O. Reichhardt, New J. Phys. **18**, 095005 (2016).

[48]	R. Gruber *et al.*, Nat. Commun. **13**, 3144 (2022).

[49]	K. Litzius *et al.*, Nat. Electron. **3**, 30 (2020).

[50] X. Zhang, N. Vernier, W. Zhao, H. Yu, L. Vila, Y. Zhang, and D. Ravelosona, Phys. Rev. Appl. **9**, 024032 (2018).

[51] S. Lemerle, J. Ferré, C. Chappert, V. Mathet, T. Giamarchi, and P. Le Doussal, Phys. Rev. Lett. **80**, 849 (1998).

[52] H. Yang, M. Chshiev, B. Dieny, J. Lee, A. Manchon, and K. Shin, Phys. Rev. B **84**, 054401 (2011).

[53] B. Dieny and M. Chshiev, Rev. Mod. Phys. **89**, 025008 (2017).

[54] N. Nagaosa, J. Sinova, S. Onoda, A. H. MacDonald, and N. P. Ong, Rev. Mod. Phys. **82**, 1539 (2010).

[55] L. Liu, C.-F. Pai, Y. Li, H. Tseng, D. Ralph, and R. Buhrman, Science **336**, 555 (2012).

[56] Q. Hao and G. Xiao, Phys. Rev. B **91**, 224413 (2015).

[57] K. Wang, L. Qian, W. Chen, S.-C. Ying, G. Xiao, and X. Wu, Phys. Rev. B **99**, 184410 (2019).

[58] X. Wang, H. Yuan, and X. Wang, Commun. Phys. **1**, 31 (2018).

[59] D. Maccariello, W. Legrand, N. Reyren, K. Garcia, K. Bouzehouane, S. Collin, V. Cros, and A. Fert, Nat. Nanotechnol. **13**, 233 (2018).

[60] L. Desplat and J.-V. Kim, Phys. Rev. Lett. **125**, 107201 (2020).

[61] L. Desplat, C. Vogler, J.-V. Kim, R. L. Stamps, and D. Suess, Phys. Rev. B **101**, 060403 (2020).

[62] L. Desplat and J.-V. Kim, Phys. Rev. Appl. **14**, 064064 (2020).

**Supplementary Materials**

**Electronic Noise of a Single Skyrmion**

Kang Wang,* Yiou Zhang, Vineetha Bheemarasetty, See-Chen Ying and Gang Xiao†

**Affiliation**

*Department of Physics, Brown University, Providence, Rhode Island 02912, USA*

---

* kang_wang@brown.edu
† gang_xiao@brown.edu

**Contents**

**Supplementary Note I – Sample fabrication**

We deposit the multilayer samples by magnetron sputtering on thermally oxidized silicon wafers with a base pressure of $5.0 \times 10^{-8}$ Torr. The multilayers are Ta (2.3 nm, $P_{Ta} = 3$ W)/Co$_{40}$Fe$_{40}$B$_{20}$ (0.95 nm)/MgO (1.6 nm)/TaO$_x$ (1.0 nm), Ta (2.0 nm, $P_{Ta} = 4$ W)/Co$_{40}$Fe$_{40}$B$_{20}$ (0.95 nm)/MgO (1.6 nm)/TaO$_x$ (2.0 nm), and Ta (2.0 nm, $P_{Ta} = 5$ W)/Co$_{40}$Fe$_{40}$B$_{20}$ (0.95 nm)/MgO (1.6 nm)/TaO$_x$ (2.0 nm), corresponding to the strong, intermediate, and weakly pinned samples, respectively. The MgO layer is deposited using RF power with an argon pressure of 0.7 mTorr while other layers are deposited using DC power with an argon pressure of 1.4 mTorr. We use 15-Watt DC power for the deposition of the ferromagnetic Co$_{40}$Fe$_{40}$B$_{20}$ layer and a lower DC power ($P_{Ta} = 3, 4, 5$ Watt) for the deposition of the Ta layer. The Ta DC power $P_{Ta}$ is regulated to implement moderate pinning strengths [see Supplementary Note V for details]. The geometries of the Hall-cross are constructed using photolithography and physical ion milling. The samples are then annealed at different temperatures in a high-vacuum chamber with a vacuum pressure of $1.0 \times 10^{-6}$ Torr in the presence of a magnetic field of about 0.4 T normal to the sample plane.

**Supplementary Note II – Experimental setup**

We study skyrmions by measuring magnetically induced electronic noise with the help of direct imaging using a homemade polar magneto-optic Kerr effect (p-MOKE) microscope. Supplementary Figure 1(a) presents a schematic of the experimental setup. The sample is placed on a printed circuit board (PCB) and is connected to the PCB using a wire-bonding machine (HB10, TPT), which is further connected to external circuits as shown in Supplementary Fig. 1(b). We use a polyimide flexible heater placed at the bottom of the PCB to heat the sample. In the PCB, we also introduce a thermal via, which is a hole that connects the top and bottom surfaces to enhance the thermal conductivity between the heater and the sample. We place an electromagnet under the PCB to supply a perpendicular magnetic field. The temperature and the magnetic field are detected via a temperature sensor and a Hall sensor, respectively, which are both placed close to the sample on the PCB. We wait for a sufficiently long time before collecting measurements to ensure the sample temperature is stable.

We detect electronic noise signals by measuring the resultant Hall voltages when a DC current is applied through a Keithley 6221 DC current source. The Hall voltages are intensified through a pair of amplifiers (INA828, Texas Instruments) and the amplified signals from these two independent amplifiers are then converted to digital signals through a high-speed data acquisition (DAQ) device (USB-1602HS, Measurement Computing) with 16-bit resolution and a high sample rate of 2 MS/s. We note from the source specifications (see this link) that the current source has a peak-to-peak noise of 200 nA at 0.1 Hz and 40 nA at 10 Hz when the output current is in the range of 2 mA. This noise is significantly lower than the Hall-resistance noise in our studies. The Hall voltage is given by $(V_H + \delta V_H) = (R_H + \delta R_H)(I + \delta I)$ where $\delta R_H$ and $\delta I$ are the Hall-resistance and current variations, respectively. From this, we obtain the Hall-voltage noise $\frac{\delta V_H}{V_H} = \frac{\delta R_H}{R_H} + \frac{\delta I}{I} + \frac{\delta R_H}{R_H}\frac{\delta I}{I} \cong \frac{\delta R_H}{R_H}$ and thereby $\delta R_H \approx \delta V_H/I$ considering $\frac{\delta I}{I} \ll \frac{\delta R_H}{R_H}$. We also assume a negligible contribution from the current noise using the fact that the noise at the saturated ferromagnetic state is much lower than the noise of other magnetic structures. Noise from the amplifiers is cancelled out through a cross-correlation method [1-4]. This allows us to collect accurate measurements of the electronic noise signals from a single skyrmion.

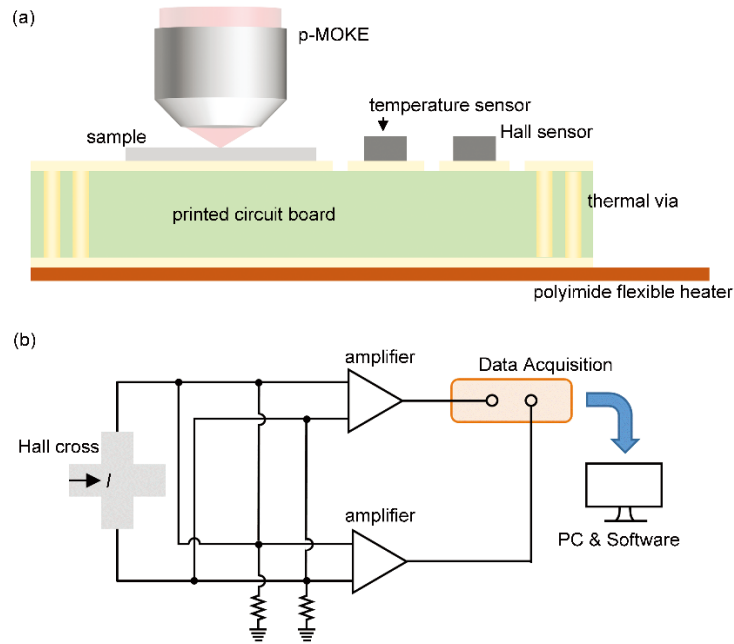

Supplementary FIG. 1. Experimental setup. (a) Schematic of the experimental setup. (b) Circuit diagram of electronic measurements.

The p-MOKE microscope for magnetic imaging consists of a 633 nm wavelength laser and 50× objective lens with numerical aperture of 0.80. This gives rise to a maximum achievable resolution of about 0.48 µm, which is notably smaller than the size of the skyrmions in this study. We aim to achieve the highest resolution possible in p-MOKE measurements to accurately extract the skyrmion profile, as shown in the inset of Fig. 3(b), directly from the CCD camera recording. To determine the average skyrmion diameter, we measure multiple skyrmions in p-MOKE images using the Hall cross as a reference length scale.

The variation of skyrmion size with $H_z$ is well fitted by a theorical model [5] with adopting the effective PMA $K_{u,eff} = 9200$ J/m$^3$, saturation magnetization $M_S = 1.06$ MA/m, exchange stiffness $A = 15$ pJ/m, and the DMI $D_{Int} = 0.193$ mJ/m$^2$. The $M_S$, $A$ and $D_{Int}$ parameters reflect our previous measurements [6]. We note that the DMI size supports the formation of Néel-type domain walls [6]. We also show more evidence of the Néel-type configuration in our previous studies [6,7].

**Supplementary Note III – Magnetic properties of the magnetic multilayer**

Magnetic configurations in magnetic multilayers are determined by the competition between multiple energy terms including the perpendicular magnetic anisotropy (PMA) ($U_{ani}$), exchange interaction ($U_{ex}$), interfacial Dzyaloshinskii-Moriya interaction (DMI) ($U_{DMI}$), Zeeman ($U_{Zeeman}$) and demagnetization ($U_{demag}$) energies

$$\begin{aligned} U &= U_{ani} + U_{ex} + U_{DMI} + U_{Zeeman} + U_{demag} \\ &= -\int d^2x K_{u,eff} m_z^2 + \int d^2x A(\nabla \boldsymbol{m})^2 \\ &\quad + \int d^2x D_{Int}[m_z \text{div}\, \boldsymbol{m} - (\boldsymbol{m} \cdot \nabla) m_z] - \mu_0 H_z \int d^2x M_S m_z \\ &\quad - \frac{1}{2}\mu_0 \int d^2x M_S \boldsymbol{m} \cdot \boldsymbol{H}_{demag}. \end{aligned} \quad (S1)$$

In Eq. S1, $\boldsymbol{m} = (m_x, m_y, m_z)$ is the normalized magnetization, $H_z$ is the perpendicular magnetic field, $M_S$ is the saturation magnetization, $A$ is the exchange stiffness, $K_{u,eff}$ is the effective PMA, $D_{Int}$ is the DMI constant and $\boldsymbol{H}_{demag}$ is the demagnetization field. Here, we make a valid two-dimensional approximation of the magnetic system as the lateral dimensions of the Hall cross ($20 \times 20$ μm$^2$ and $5 \times 5$ μm$^2$), skyrmions (on the order of 1 μm), and the domain-wall width (on the order of 10 nm) are all much larger than the ferromagnetic layer thickness (0.95 nm).

Supplementary Figure 2 displays Hall resistance $R_H$ measurements as a function of a perpendicular magnetic field $H_z$ for the strongly pinned magnetic thin film ($P_{Ta} = 3$ Watt) that was annealed at different temperatures $T_a$. The Hall resistance is dominated by the anomalous Hall resistance which is proportional to the perpendicular magnetization while the ordinary Hall resistance is negligible in comparison [8]. Square loops with sharp magnetization reversal are observed at intermediate annealing temperatures (135 °C and 150 °C). This is associated with a strong PMA and a uniform ferromagnetic state. With increasing or decreasing the $T_a$, the hysteresis loops slant with a small remanence that suggests a weak PMA and a magnetic multidomain state at zero magnetic field [100 °C, see details in

Supplementary Fig. 3(a)]. When the $T_a$ is further increased or decreased, the loops slant with zero remanence and exhibit large saturation fields (as grown and 300 °C). This corresponds to an in-plane magnetization state. In this work, we annealed our samples at different temperatures to obtain moderate PMA. For our study, we chose magnetic thin films that were annealed at 100 °C, 297 °C, and 105 °C, for the strong, intermediate, and weakly pinned samples, respectively, where skyrmions can be created in a suitable field range.

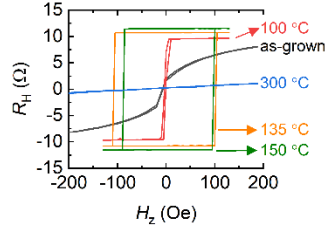

Supplementary FIG. 2. Annealing effect on magnetic properties of a magnetic thin film. Hall resistance $R_H$ measurement results as a function of the perpendicular magnetic field $H_z$ for the strongly pinned magnetic thin film that was annealed at different temperatures.

The magnetic multilayers have a PMA that originates from the electronic hybridization between the $2p$ electrons of oxygen atoms in the MgO layer and the $3d$ electrons of transition metals in the magnetic layer with a spin-orbit coupling [9]. The PMA in a magnetic thin film is strongly affected by the interfacial oxidation state [10]. Either over oxidation or under oxidation weakens the interfacial PMA. Roadmacq et al. has thoroughly studied the influence of thermal annealing on the PMA in a magnetic thin film of Pt/Co/AlO$_x$ [11]. The results are similar to what we obtain in Supplementary Fig. 2. The annealing temperature affects the oxygen migration as well as the Co-O bond at the interface.

Supplementary Figure 3(a) displays the Hall-resistance $R_H$ measurement result for the strongly pinned sample under study in the main text. In addition to the PMA, we also observe an exchange bias which may originate from the Co$_{40}$Fe$_{40}$B$_{20}$/MgO interface [12,13]. At the middle field $H_z = -2.90$ Oe, a zero Hall resistance is observed, which indicates an absence of the remnant magnetization. At this field, we observe through the p-MOKE microscope a magnetic multidomain state [Supplementary Fig. 3(a)]. Increasing or decreasing the field leads to an increase or decrease in the Hall resistance, respectively, and leads magnetic multidomains to transition into skyrmions with reversed polarities [Supplementary Fig. 3(a)]. Magnetic skyrmions are stabilized by the interfacial DMI which originates from the manipulation of exchange interactions between two atoms of the ferromagnetic Co$_{40}$Fe$_{40}$B$_{20}$ layer by a third nonmagnetic atom in the Ta layer with large spin-orbit coupling [14,15].

We also show Hall-resistance $R_H$ measurements and p-MOKE images for the weakly pinned sample in Supplementary Fig. 3(b). The Hall curve shows a negligible coercivity in comparison to that of the magnetic multilayer in the strong pinning regime.

In the weakly pinned sample, both the saturation field and the saturated Hall resistance are smaller [Supplementary Fig. 3]. This observation suggests a larger perpendicular magnetic anisotropy (PMA) and a smaller saturation magnetization in the weakly pinned sample. The different saturated Hall resistance may be a result of the different layer thicknesses of the magnetic thin films as discussed in Supplementary Note I]. The difference in the Hall signals does not affect our main conclusion of the distinct types of noise signatures arising from thermally induced and current-induced dynamics of a skyrmion in different pinning regimes.

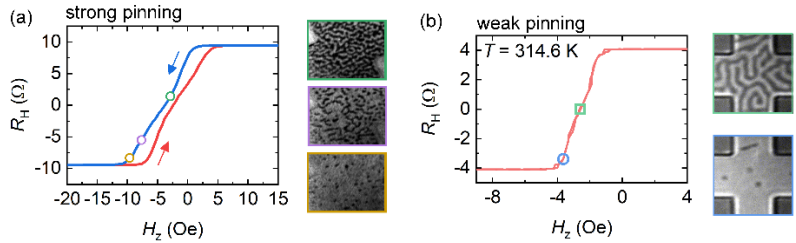

Supplementary FIG. 3. Magnetic properties of differently pinned samples. (a) Hall resistance $R_H$ measurement result and p-MOKE images as a function of the perpendicular magnetic field $H_z$ for the strongly pinned magnetic thin film. (b) Hall resistance $R_H$ measurement result and p-MOKE images as a function of the perpendicular magnetic field $H_z$ for the magnetic multilayer in the weakly pinned magnetic thin film.

**Supplementary Note IV – The evidence of skyrmions and their manipulations**

Magnetic skyrmions are topologically protected quasi-particles with non-zero topological charges $Q = \pm 1$. Due to its topology, the skyrmion experiences a skyrmion Hall effect under a driving force. On biasing with a current, we observe that the magnetic bubbles in the samples acquire an additional transverse velocity component in addition to the longitudinal motion along the current direction. Moreover, the transverse component reverses when reversing the polarity of the magnetic bubbles. The observation of the skyrmion Hall effect confirms that the magnetic bubbles in this study are indeed skyrmions with non-zero topological charges. More descriptions of the demonstration can be found in our previous work [6,7].

The labyrinthine domain phase can transition into a multiple-skyrmion state either by increasing or decreasing the field [Supplementary Fig. 3]. In the strong and intermediate pinning regimes at a constant field near the saturation limit, various states with different numbers of skyrmions can be accessed in the Hall cross. We start from a state of multiple skyrmions at a particular magnetic field. A state with a reduced skyrmion number can be achieved by field cycling in which the field is first increased to remove some skyrmions and then decreased to stabilize this smaller number of skyrmions. Studies of isolated skyrmions at a constant field allow us to disregard the field's effect on the noise analysis.

**Supplementary Note V – Effects of the sputtering rate on the pinning of magnetic structures**

We regulate the DC power ($P_{Ta}$) for deposition of the Ta layer to implement moderate pinning strengths. We identify the three growth scenarios with $P_{Ta} = 3, 4,$ and 5 Watt as corresponding to the strong, intermediate, and weakly pinned samples, respectively. We observe through a p-MOKE microscope that the domain walls and skyrmions, under a constant current $I = 0.5$ mA, move negligibly, creep forward [Supplementary Movie 1], and flow steadily [Supplementary Movie 2] in films grown with $P_{Ta} = 3, 4,$ and 5 Watt, respectively. Moreover, we observe that skyrmions in films grown with $P_{Ta} = 5$ Watt can move freely under thermal effects [Supplementary Movie 3]. We extract trajectories of the skyrmion center from Supplementary Movie 3. As shown in Supplementary Fig. 4, the skyrmion experiences diffusive motion owing to the thermal perturbations but may get pinned in some specific sites. The movies have a low temporal resolution of 0.61 s. This makes Supplementary Movie 3 look sporadic and makes Supplementary Movie 2 blurry owing to a fast domain-wall motion.

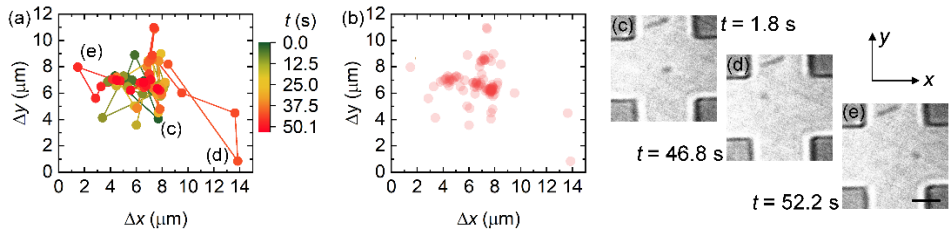

Supplementary FIG. 4. Trajectories of a skyrmion in a weakly pinned sample. (a) Trajectories of the skyrmion center extracted from Supplementary Movie 3. (b) Scatter plot of the skyrmion center position in (a). (c) – (e) P-MOKE images of the skyrmion at selected times of (c) $t = 1.8$ s, (d) $t = 46.8$ s, and (e) $t = 52.2$ s. The scale bar is 10 μm.

We note that the white noise diminishes as the pinning strength weakens, as shown in Supplementary Fig. 5. This trend indicates that more defects exist in a stronger pinning sample that serve to pin skyrmions.

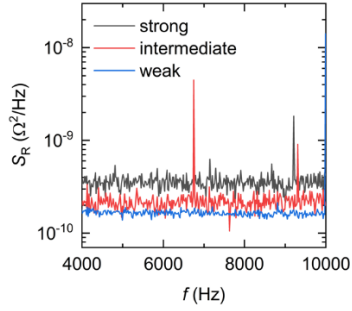

Supplementary FIG. 5. White noise in the three samples with varying pinning conditions.

We then discuss how the $P_{Ta}$ affects the thin-film properties along with other possible factors that may influence the pinning conditions of the three magnetic thin films in our study.

According to our previous studies [16] and other works [17,18] on similar multilayers grown by magnetron sputtering, one expects that the magnetic films under study possess polycrystalline or amorphous-like disordered structures. To characterize thin-film structures and the effects of $P_{Ta}$ on thin-film properties, we have grown substrate/Ta (~20 nm) films using different DC powers ($P_{Ta} = 3, 4,$ and 5 Watt) for deposition. Supplementary Figures 6(a) and (b) present x-ray diffraction (XRD) patterns of these films. Results show one peak at approximately $2\theta = 33.5°$ corresponding to the characteristic (002) peak of the β-phase Ta [19-21]. The peak position shifts gradually to lower $2\theta$ angles when the $P_{Ta}$ is increased [Supplementary Fig. 6(b)]. This suggests an expansion of the out-of-plane lattice constant $d_{002}$ [Supplementary Fig. 6(c)]. According to literature results, this may be due to the kinetic-energy ($E_{dep}$) variation of the sputtered atoms [21,22] and/or the incorporation of impurities in the films [23]. We also estimate the crystallite dimension of films by using the Scherrer equation $D = \frac{K\lambda}{\text{FWHM} \sin\theta}$ where $\lambda = 1.54$ Å is the x-ray wavelength, FWHM is the full width at half maximum of the peak, $\theta$ is the Bragg angle, and $K$ is the shape factor which is typically adopted as 0.9 [24,25]. The results in Supplementary Fig. 6(d) illustrate that the grain size increases when the $P_{Ta}$ is increased. This is again consistent with the literature results that the FWHM decreases when the $E_{dep}$ is increased [22].

Ta is a known getter material [26-28] and can thereby react with various gases [26]. It has been shown that the presence of Ta in the reaction space helps to yield very smooth and high-quality films owing to a reduction in the amount of impurities made possible by the getting properties of Ta [27,28]. Moreover, much research has shown that the Ta properties are strongly affected by growth conditions [19,21,22,29-31]. For example, Feinstein et al. [30] and Senkevich [19] have illustrated that Ta films grown onto oxygen-rich substrates are prone to form the β phase while the films grown onto substrates free of oxygen form the α phase. Knepper et al. has reported a very large change in the stress of Ta films by incorporating oxygen into the films [31]. Colin et al. [21] and Ellis et al. [22] have revealed large increases in both the out-of-plane lattice constant and crystallite dimension with increasing the $E_{dep}$.

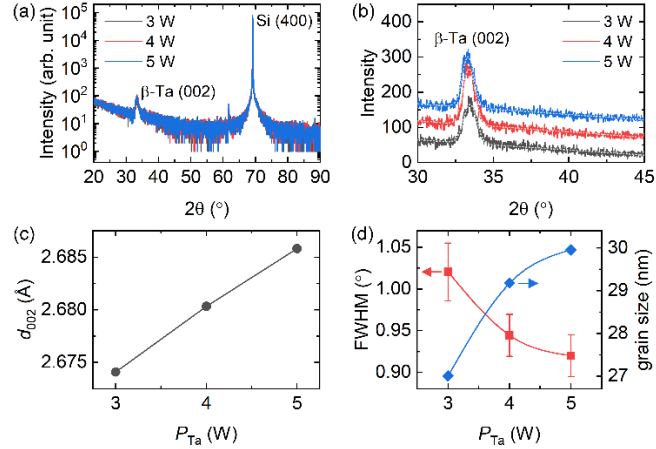

Supplementary FIG. 6. Effects of the sputter rate on material structures. (a) X-ray diffraction (XRD) patterns of substrate/Ta (approximately 20 nm) films grown by using different DC powers ($P_{Ta}$ = 3, 4, and 5 Watt) for deposition. (b) XRD patterns scanned in the range of $2\theta = 30°$ to $45°$. (c) The out-of-plane lattice constant $d_{002}$ of the β-phase Ta with various $P_{Ta}$. (d) The full width at half maximum (FWHM) of the β-Ta (002) peak and grain size in films with various $P_{Ta}$.

According to the literature results as well as our own experimental results in Supplementary Fig. 6, we assume two possible effects of the $P_{Ta}$ on thin-film structures which may lead to different pinning conditions of magnetic thin films. First, the Ta is a getter material [26-28] such that a lower sputter rate would induce more contaminants in the Ta layer. This affects the local atomic arrangements and increases spatial non-uniformities of magnetic properties. Secondly, $P_{Ta}$ may affect crystalline structures by varying the kinetic energy ($E_{dep}$) of the sputtered atoms. A lower $P_{Ta}$ (i.e., a lower $E_{dep}$ [21,22]) leads to a decrease in the out-of-plane lattice constant and a reduction in the crystallite dimension of the Ta layer [Supplementary Fig. 6]. The structural variation may also affect the pinning of magnetic thin films as well.

In addition to the structural variation, material parameters may also vary in the three samples owing to different growth conditions and film thicknesses, thereby affecting the dynamics of magnetic structures. The noise arises from the current-induced skyrmion nucleation and/or the thermally dominated internal-mode oscillation dynamics. We now discuss how material parameters affect the two dynamics respectively.

It has been well established that the energy barrier for magnetization switching is given by $\Delta E = K_{u,\text{eff}} V$ where $V$ is the size of the nucleated domain [9]. The competition between the $\Delta E$ and thermal energy $k_B T$ determines the thermal stability of the domain as well as the switching current [9]. The skyrmion size is determined by material parameters including the effective PMA $K_{u,\text{eff}}$, exchange interaction $A$, DMI $D_{\text{Int}}$, and saturation magnetization $M_S$ [32]. It has been shown that the domain width or skyrmion size decreases by decreasing the $K_{u,\text{eff}}$ [32] and $A$ [6] and/or by increasing the $D_{\text{Int}}$ [6] and $M_S$ [33]. By this, one expects a reduction in the $\Delta E$ by decreasing the $K_{u,\text{eff}}$ and A and/or by increasing the $D_{\text{Int}}$ and $M_S$ which thereby affects the current-induced domain nucleation dynamics.

The domain-wall hopping dynamics is governed by the competition between the domain-wall surface energy $E_{\text{el}}$ and the pinning energy $U_{\text{pinning}}$ which also determines the static configuration of domain walls [34-38]. Lemerle et al. has shown a field-dependent domain-wall velocity along with the critical field for domain-wall motion $H_{\text{crit}} \propto E_{\text{el}}/M_S d$ where $d$ is the ferromagnetic layer thickness [34]. The $E_{\text{el}}$ is determined by multiple parameters and increases by increasing the $A$, $K_{u,\text{eff}}$, $M_S$ and by decreasing the $D_{\text{Int}}$ [32]. This suggests that the domain wall is more strongly pinned by increasing the $A$, $K_{u,\text{eff}}$, $M_S$ and/or by decreasing the $D_{\text{Int}}$ as a larger $E_{\text{el}}$ energy variation is needed for escaping the pinning.

We confirm that the $P_{\text{Ta}}$ has a notable effect on thin-film structures which may affect the pinning of magnetic structures. Moreover, as discussed above, the material parameters may also influence the pinning of magnetic structures. Decreases in the $A$, $K_{u,\text{eff}}$, $M_S$ and/or an increase in the $D_{\text{Int}}$ weaken the pinning effect on domain structures. These parameters may vary in the three samples owing to different growth conditions and film thicknesses, thereby affecting the dynamics of magnetic structures.

**Supplementary Note VI – Electronic noise of a skyrmion in the strong pinning regime**

We have performed electronic noise measurements of isolated skyrmions for multiple samples in the strong pinning regime. The results are presented in Supplementary Fig. 7 and Supplementary Fig. 8. We obtain the average contribution of each skyrmion to the Hall-resistance noise $\Delta S_R^{\text{Skyr}}$ and the contribution from the uniform magnetization state $S_{R,0}$ for each sample. The $S_{R,0}$ is the noise of the uniform ferromagnetic state and hence suggests its electronic origin. These results are self-consistent and clearly indicate the $1/f^\gamma$ noise signature of the $\Delta S_R^{\text{Skyr}}$, although the $S_{R,0}$ may differ among samples.

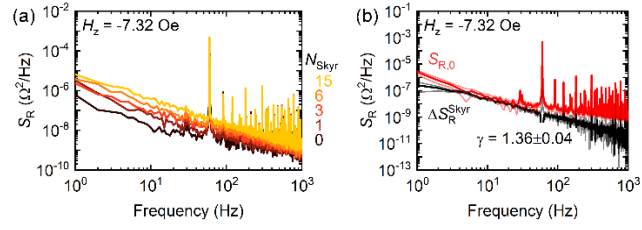

Supplementary FIG. 7. Electronic noise of states with variable skyrmion numbers in the strong pinning regime. (a) Hall-resistance noise spectra $S_R$ for states with different skyrmion numbers $N_{Skyr}$ measured at $H_z = -7.32$ Oe. (b) The noise spectra of $S_{R,0}$ and $\Delta S_R^{Skyr}$.

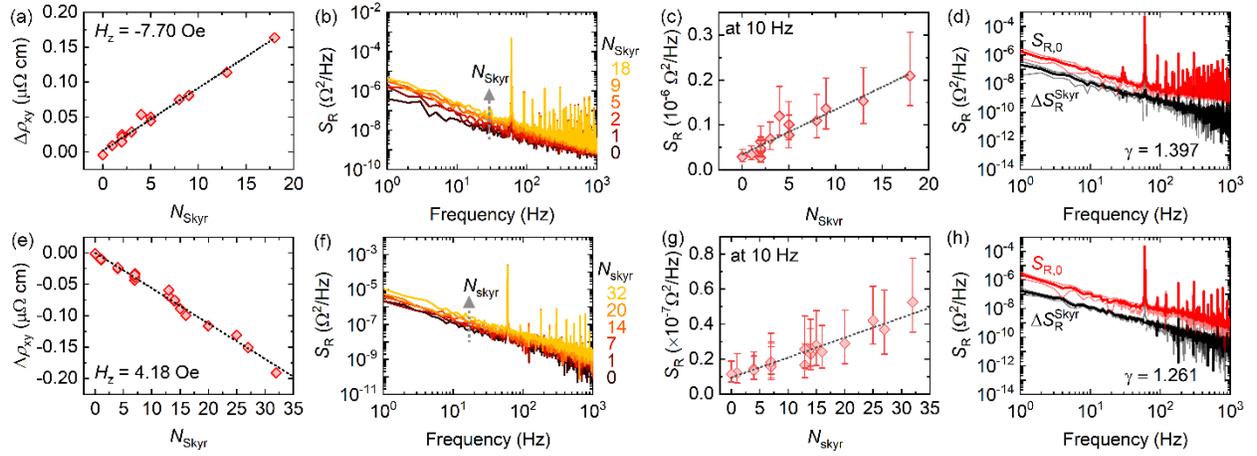

Supplementary FIG. 8. $1/f^\gamma$ noise of a skyrmion in two strongly pinned magnetic thin films. Electronic noise measurements of isolated skyrmions at fields of (a) – (d) $H_z = -7.70$ Oe and (e) – (h) $H_z = 4.18$ Oe for two different samples. (a), (e) The relative Hall resistivity $\Delta \rho_{xy}$ as a function of the skyrmion number $N_{Skyr}$ in the Hall cross. The dotted line is a linear fit. (b), (f) Hall-resistance noise spectra $S_R$ for states with different skyrmion numbers $N_{Skyr}$. (c), (g) Hall-resistance noise at 10 Hz as a function of $N_{Skyr}$. The dotted line is a fit to $S_R = S_{R,0} + \Delta S_R^{Skyr} N_{Skyr}$ where $S_{R,0}$ is the background noise and $\Delta S_R^{Skyr}$ is the average contribution of each skyrmion to the noise. (d), (h) The noise spectra of $S_{R,0}$ and $\Delta S_R^{Skyr}$.

In derivation of the electronic noise of an isolated skyrmion, we assume negligible interactions between skyrmions due to their large separations. The skyrmion-skyrmion pair interaction was first explored theoretically to illustrate an exponential decay of the repulsive force with the pair's separation [39]. It was shown that this interaction is mediated by the overlap of spin fields of the two skyrmions. In addition, neighboring skyrmions can also interact through the demagnetization field. The demagnetization field-mediated skyrmion-skyrmion interaction has been demonstrated to be important in forming a skyrmion lattice in a frustrated magnet [40] and in determining the spatial arrangement of multiple skyrmions in confined geometries [41]. We have calculated the demagnetization field that one skyrmion generates at the other skyrmion's position [42]. The demagnetization field also exhibits a rapid decay and only has a notable effect on neighboring skyrmions. Since we study isolated skyrmions with large separations between them, the skyrmion-skyrmion pair interaction can be neglected in analyzing the electronic noise of isolated skyrmions.

**Supplementary Note VII – Micromagnetic simulations of skyrmion noise in the strong pinning regime**

In the strong pinning regime, we observe the $1/f^\gamma$ noise of a skyrmion. According to van der Ziel's picture, the $1/f^\gamma$ noise signature arises from a collection of non-identical RTN oscillators. In the case of a skyrmion in the strong pinning regime, we speculate that the RTN oscillator can be attributed to the internal domain-wall hopping dynamics. To validate this speculation, we have conducted micromagnetic simulations using the finite-difference solver MUMAX3 based on the graphic processing unit [43]. The simulations involve solving the Landau-Lifshitz-Gilbert (LLG) equation to obtain the time-dependent normalized magnetization $\bm{m}$. For micromagnetic simulations, we adopted the saturation magnetization $M_S = 1.06 \times 10^6$ A/m, the exchange stiffness $A = 1.5 \times 10^{-11}$ J/m, the averaged PMA $K_u = 7.58 \times 10^5$ J/m$^3$, the averaged DMI constant $D_{Int} = 0.193 \times 10^{-3}$ J/m$^2$, and the Gilbert damping constant $\alpha = 0.1$. These magnetic parameters were determined based on previous measurements [6] and can accurately describe the magnetic materials used in our study due to their similar multilayer structure and growth conditions. To simulate the presence of pinning centers, we introduced grains with an average size of 10 nm and random distributions of interfacial PMA and DMI. We set the random PMA variation at $\frac{\Delta K_u}{K_u} = 7\%$ and the random DMI variation at $\frac{\Delta D_{Int}}{D_{Int}} = 21\%$. Additionally, to account for thermal fluctuations at 300 K, we included a stochastic thermal field in the LLG equation. The simulation results are presented in Supplementary Fig. 9.

Supplementary Fig. 9(a) depicts the skyrmion configuration, and we have highlighted the skyrmion boundaries at three different times using solid dots. In Supplementary Fig. 9(b), we present the perpendicular magnetization $m_z$ in time in three representative regions that are marked in Supplementary Fig. 9(a). The simulation results shown in Supplementary Figs. 9(a) and (b) clearly demonstrate the occurrence of multiple internal domain-wall hopping oscillations within the skyrmion. It is noteworthy that a collection of these internal domain-wall hopping dynamics gives rise to a $1/f^\gamma$ noise, as illustrated in Supplementary Fig. 9(c). The simulation results well reproduce the experimental observation of the $1/f^\gamma$ noise in a skyrmion within the strong pinning regime.

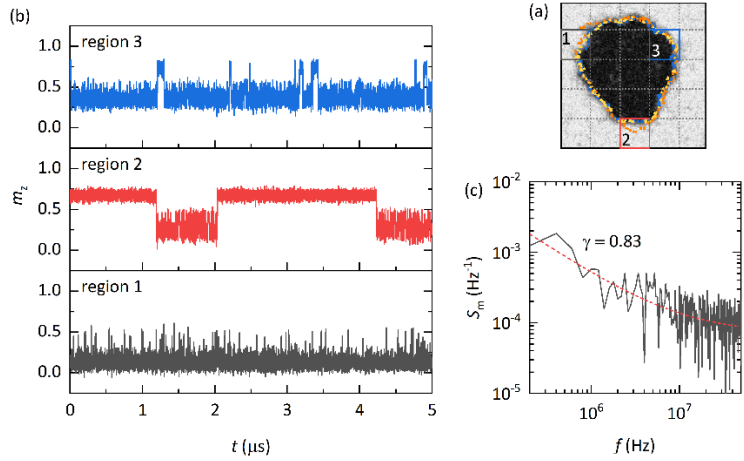

Supplementary FIG. 9. (a) Map of the perpendicular magnetization $m_z$ for the skyrmion used in micromagnetic simulations. The black region corresponds to $m_z = -1$, while the white region represents $m_z = 1$. We highlight the skyrmion boundaries at three different times using solid dots. (b) Temporal variations of $m_z$ in three distinctive representative regions marked in panel (a). (c) The power spectrum of $m_z$ for the entire region in panel (a). The data is fitted with a $1/f^\gamma$ noise model in addition to a white noise component, as denoted by the dashed line.

## Supplementary Note VIII – Estimate of the strength of pinning

The pinning energy $U_\text{pinning}$, in conjunction with the thermal energy $k_B T$, plays a crucial role in governing the internal domain-wall hopping dynamics through the Arrhenius law $f_0 = f_{00}\exp(-U_\text{pinning}/k_B T)$, where $f_{00}$ is the attempt frequency. Consequently, the $U_\text{pinning}$ distribution affects the distribution of the internal domain-wall hopping oscillators as well as the noise signature from this collection of non-identical oscillators. Conversely, one can deduce the distribution of $U_\text{pinning}$ by analyzing the noise spectrum.

In derivation of the $U_\text{pinning}$ distribution, we assume a constant attempt frequency $f_{00} = 1$ GHz. The $U_\text{pinning}$ is then derived using $U_\text{pinning} = -k_B T \ln(f_0/f_{00})$. Currently, no single solid function exists to precisely describe the $U_\text{pinning}$ distribution [35,44-46]. We assume that the $U_\text{pinning}$ distribution follows a log-normal distribution $P(U_\text{pinning}) = \frac{1}{U_\text{pinning}\sigma\sqrt{2\pi}}\exp\left(-\frac{(\ln U_\text{pinning}-\mu)^2}{2\sigma^2}\right)$. The $U_\text{pinning}$ distribution $P(U_\text{pinning})$ is then derived by fitting the noise spectrum with $S_R(f) = \int_0^{+\infty} P(U_\text{pinning}(f_0)) \frac{A}{2f_0} \frac{1}{(1+(\pi f/f_0)^2)} df_0$ where $\sigma$, $\mu$, and $A$ are fitting parameters. Figure 3(a) in the main text displays the $U_{\text{pinning}}$ distribution contributing to the $1/f^\gamma$ noise of a skyrmion in the strong pinning regime. This provides valuable insights into the strength of pinning in a magnetic system.

**Supplementary Note IX – Electronic noise of magnetic domain walls in the strong pinning regime**

We also study the electronic noise of magnetic multidomains as a comparison. The results, as presented in Supplementary Figs. 10(a) and (b), demonstrate that the Hall-resistance noise decreases with increasing the domain width $W$. However, it is noteworthy that the effect of domain-size variation on the electronic-noise variation is much smaller in comparison to that of skyrmions [Fig. 3(e) in the main text]. The dependence of the noise on domain size in Supplementary Figs. 10(a) and (b) aligns with the common expectation that the number of internal-mode oscillators is proportional to the total length of domain walls, consequently linearly affecting the noise amplitude.

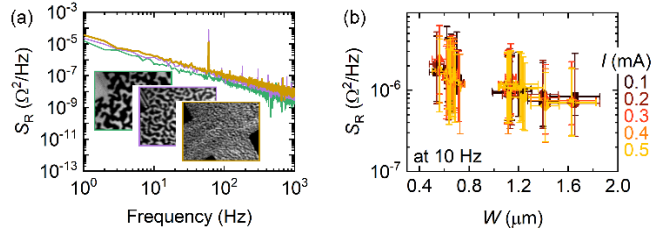

Supplementary FIG. 10. Electronic noise of multidomains. (a) Hall-resistance noise spectra of magnetic multidomains (inset) with different domain widths $W$. The lines are in the same colors of corresponding rectangles of the p-MOKE images. (b) Hall-resistance noise at 10 Hz as a function of the domain width.

**Supplementary Note X – Electronic noise of skyrmions in the intermediate pinning regime**

In Fig. 4 in the main text, the skyrmion is strongly anchored such that we can detect the telegraph noise resulting from its local size fluctuations. When a large current is applied, the skyrmion escapes from the local pinning center and undergoes the skyrmion Hall effect, as illustrated in Supplementary Fig. 11. It moves across weak pinning sites before being pinned again on another strong pinning center.

We observe the random telegraph noise (RTN) signature for multiple isolated skyrmions [Supplementary Fig. 12, also see Supplementary Fig. 6 in the Ref. [42]). As discussed in the main text, the RTN arises from thermally induced hopping of the mobile part of a skyrmion between two pinning sites [42]. This shares the same physics as the internal domain-wall hopping oscillations in the strong pinning regime and could lead us to a better understanding of the magnetization dynamics and the emergent electronic noise characteristics.

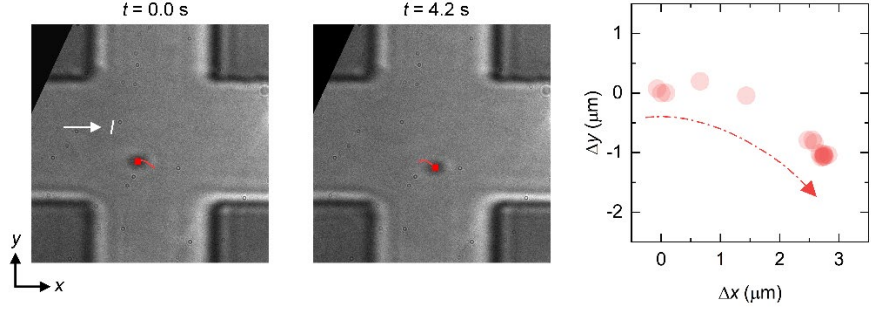

Supplementary FIG. 11. The dynamic motion of a skyrmion in the intermediate pinning regime. Under a driving current of $I = 0.5$ mA, a skyrmion escapes from the local pinning center and undergoes the skyrmion Hall effect. It moves across weak pinning sites before being pinned again on another strong pinning center. Lines are trajectories of the skyrmion under this current. Symbols are positions of the skyrmion at the corresponding time of each image. The panel on the right displays a scatter plot depicting the position of the skyrmion center over time.

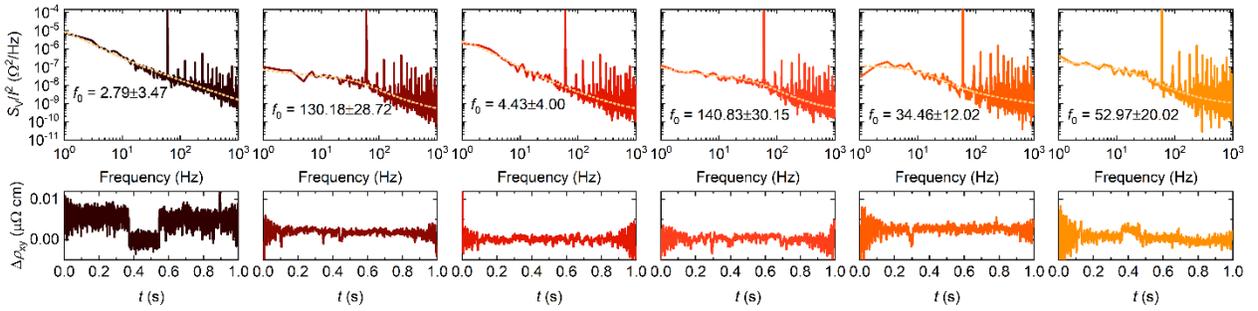

Supplementary FIG. 12. Random telegraph noise of multiple isolated skyrmions in the intermediate pinning regime. Hall-resistance noise $S_R$ spectra and the relative Hall resistivity variation $\Delta\rho_{xy}$ of multiple isolated skyrmions in the intermediate pinning regime over time. Dotted lines are fits to Equation 1 in the main text which yield the fluctuation rate $f_0$ of the RTN.

We measure the electronic noise of states with variable skyrmion numbers $N_{Skyr}$ in the Hall cross as shown in Supplementary Fig. 13. It is evident that the RTN transitions into a $1/f^\gamma$ noise signature by increasing the skyrmion number $N_{Skyr}$. The $1/f^\gamma$ signature is observed even for the state of only 5 skyrmions. Moreover, the sum of any 5 noise spectra for isolated skyrmions shown in Supplementary Fig. 12 also produces the $1/f^\gamma$ signature with $\gamma$ in the range between 1.10 and 1.15. This may provide evidence that van der Ziel's picture of the $1/f^\gamma$ noise spectrum arising from a collection of non-identical RTN oscillators exists naturally in a skyrmion-based system. The variation of $\Delta V_H^2/2I^2$ with $f_0$ [Fig. 4(f) in the main text], together with the distribution of the RTN oscillators [Fig. 3(a) in the main text, larger distributions of the RTN oscillators with higher $U_{\text{pinning}}$ and thereby lower $f_0$], explains the $1/f^\gamma$ noise of a skyrmion with $\gamma > 1$ in the strong pinning regime.

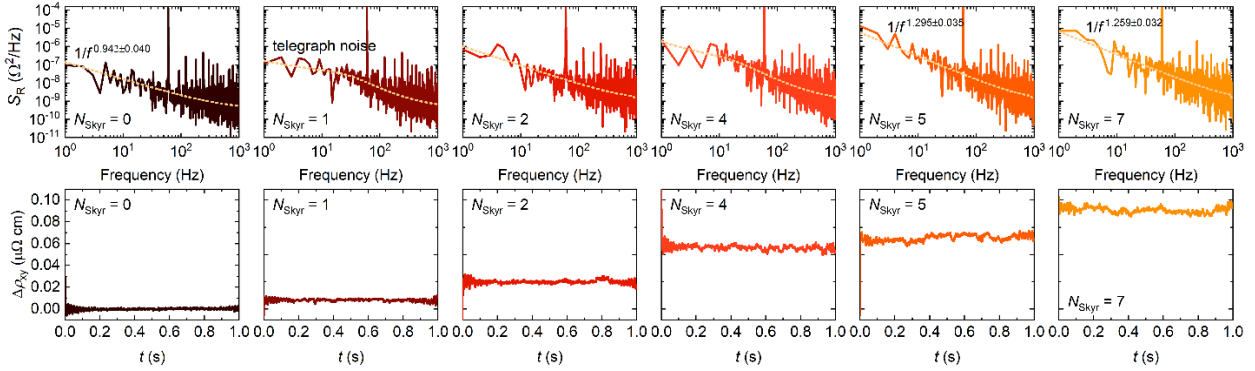

Supplementary FIG. 13. Electronic noise of states with variable skyrmion numbers in the intermediate pinning regime. Hall-resistance noise $S_R$ spectra and the relative Hall resistivity variation $\Delta\rho_{xy}$ of states with variable skyrmion numbers $N_{\text{Skyr}}$ in the Hall cross as a function of time.

**Supplementary Note XI – Temperature variation due to Joule heating**

To estimate the temperature variation due to Joule heating, we measure the longitudinal resistance $R_{xx}$ as a function of temperature [Supplementary Fig. 14(a)]. We also measure the longitudinal resistance by varying the amplitude of the current [Supplementary Fig. 14(b)]. From the resistance variation, we estimate the temperature variation due to Joule heating. Over the range of electrical currents, we apply in this work [Supplementary Fig. 14(c)], the temperature variation is smaller than 1 K, which is negligible in comparison to the range of temperatures we study.

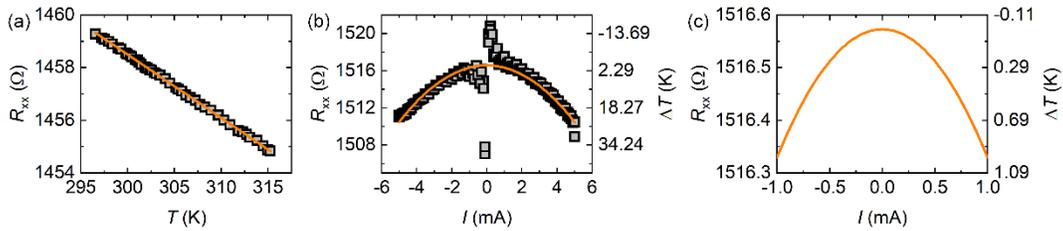

Supplementary FIG. 14. Temperature variation due to Joule heating. (a) The longitudinal resistance $R_{xx}$ as a function of the temperature. The average variation of $R_{xx}$ is around $-0.01651\%/K$. Dots are experimental results, and the orange line is a linear fit. (b) The longitudinal resistance $R_{xx}$ as a function of current. From the resistance variation, the temperature variation $\Delta T$ due to Joule heating can be estimated.

Gray squares are experimental results, and the orange line is a parabolic fit. (c) The enlarged figure of (b) in the current range that we study in this work.

**Supplementary Note XII – Discussions on other structures in physics**

Noise measurements have been used to study many other condensed matter states including superconducting vortices [47-50] and sliding charge density waves [51-53]. These structures share similar fundamental behaviors to skyrmions. For example, Marley et al. has studied the low-frequency noise of superconducting vortices and reported an increase in the noise followed by a decrease with increasing the driving force [49]. The noise reaches a maximum in the plastic flow regime. Correspondingly, the $\gamma$ value of the $1/f^\gamma$ noise increases from approximately 1 to approximately 2 and then decreases to approximately 0 with increasing the driving force. Marley et al. attributed the low noise and the $\gamma$ value approaching zero in the fluid flow regime to an incoherent addition of fluctuations from many channels. Note that we observe a similar dependency of the low-frequency noise on the driving current [Supplementary Fig. 15]. Moreover, if we fit the low-frequency noise by the $1/f^\gamma$ spectrum, the $\gamma$ value would also approach zero in the fluid flow regime at a large driving current owing to the relatively flat noise in the low-frequency regime while it is the RTN in the view of a broader frequency range. The low-frequency noise in the fluid regime decreases with current because both the current-induced skyrmion nucleation and the internal-mode oscillation dynamics shift to a higher frequency. In addition, Yoshida et al. has also observed the transition from the $1/f^\gamma$ noise to the RTN with increasing the bias current in the motion of charge density waves [53]. A complete picture connecting the noise and the dynamics of these condensed matter states, particularly a single state, however, is still missing. Our results provide a complete picture connecting the dynamics of isolated magnetic structures and their electronic noise. In our study of a skyrmion, the noise produces a $1/f^\gamma$ signature in the strong pinning regime, but a random telegraph noise in the intermediate pinning regime. Two telegraph-like signals are detected in the weak pinning regime. Distinct signatures arise from the thermally dominated internal-mode oscillation dynamics and/or the current-induced domain nucleation accompanying the statistical spatial average with distinct pinning conditions. This may help us to better understand other types of structures in physics that share similar fundamental behaviors.

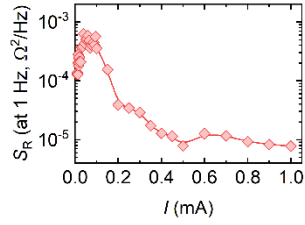

Supplementary FIG. 15. Current effect on electronic noise of skyrmions. Hall-resistance noise $S_R$ of skyrmions as a function of current in a weakly pinned sample. The results are measured at 319.5 K and $H_z = -3.58$ Oe. 1 mA current corresponds to the current density of $1.2 \times 10^{10}$ A/m$^2$ flowing in the Ta buffer layer.

**Supplementary Movie 1**

**Creep motion of magnetization structures in the intermediate pinning regime.**

This movie is recorded by a p-MOKE microscope on the magnetic film grown using DC power $P_{Ta} = 4$ Watt for the deposition of the Ta layer. This movie is acquired at a perpendicular magnetic field $H_z = -5.05$ Oe, at a constant current $I = 0.5$ mA and at the temperature $T = 307.5$ K. This current corresponds to the current density of $6.1 \times 10^9$ A/m² flowing in the Ta buffer layer. The temporal resolution of the movie is 0.61 s.

**Supplementary Movie 2**

**Steady flow of magnetization structures in the weak pinning regime.**

This movie is recorded by a p-MOKE microscope on the magnetic film grown using DC power $P_{Ta} = 5$ Watt for the deposition of the Ta layer. This movie is acquired at a perpendicular magnetic field $H_z = -2.65$ Oe, at a constant current $I = 0.5$ mA and at the temperature $T = 305.8$ K. This current corresponds to the current density of $6.1 \times 10^9$ A/m² flowing in the Ta buffer layer. The temporal resolution of the movie is 0.61 s.

**Supplementary Movie 3**

**Brownian motion of a skyrmion in the weak pinning regime.**

This movie is recorded by a p-MOKE microscope on the magnetic film grown using DC power $P_{Ta} = 5$ Watt for the deposition of the Ta layer. This movie is recorded at zero applied current, $H_z = -3.52$ Oe, and $T = 312.7$ K. The temporal resolution of the movie is 0.61 s. Brownian motion of magnetic multidomains under microscopic thermal fluctuations is observed.

**Supplementary Movie 4**

**Current-induced motion of a skyrmion in the intermediate pinning regime.**

This movie is recorded by a p-MOKE microscope on the magnetic film grown using DC power $P_{Ta} = 4$ Watt for the deposition of the Ta layer. This movie is recorded at $H_z = -5.03$ Oe, at a constant current $I = -0.5$ mA and the temperature $T = 307.1$ K. This current corresponds to the current density of $6.1 \times 10^9$ A/m² flowing in the Ta buffer layer. The temporal resolution of the movie is 0.61 s.

**Supplementary Movie 5**

**RTN of a single skyrmion in the intermediate pinning regime.**

This movie is recorded by a p-MOKE microscope on the magnetic film grown using DC power $P_{Ta} = 4$ Watt for the deposition of the Ta layer. This movie is recorded at zero applied current, $H_z = -5.76$ Oe, at a constant current $I = -0.2$ mA and the temperature $T = 307.1$ K. The temporal resolution of the movie is 0.61 s. This current corresponds to the current density of $2.4 \times 10^9$ A/m² flowing in the Ta buffer layer. The fluctuation of a skyrmion in time between the small-skyrmion and large-skyrmion configuration is observed.

**Supplementary Movie 6**

**Commensurability effect in the weak pinning regime.**

This movie is recorded by a p-MOKE microscope on the magnetic film grown using DC power $P_{Ta} = 5$ Watt for the deposition of the Ta layer. This movie is recorded at zero applied current, $H_z = -3.67$ Oe, and $T$=318.2 K. The temporal resolution of the movie is 0.61 s.

**Supplementary Movie 7**

**Current-induced nucleation and motion of isolated domains in the weak pinning regime**.

This movie is recorded by a p-MOKE microscope on the magnetic film grown using DC power $P_{Ta} = 5$ Watt for the deposition of the Ta layer. This movie is recorded at a small current of $I$=0.07 mA, $H_z = -3.16$ Oe, and $T$=318.3 K. This current corresponds to the current density of $3.4 \times 10^9$ A/m² flowing in the Ta buffer layer. The temporal resolution of the movie is 0.61 s. At this current, a sequence of isolated domains are nucleated and propagate through the Hall cross, leading to the Hall-voltage variation in time between two discrete values as shown in the inset in Fig. 6B in the main text.

**Supplementary References**

[1] M. Sampietro, L. Fasoli, and G. Ferrari, Rev. Sci. Instrum. **70**, 2520 (1999).

[2] K. Wang, Y. Zhang, and G. Xiao, Phys. Rev. Appl. **13**, 064009 (2020).

[3] Y. Zhang, K. Wang, and G. Xiao, Appl. Phys. Lett. **116**, 212404 (2020).

[4] K. Wang, Y. Zhang, S. Zhou, and G. Xiao, Nanomaterials **11**, 854 (2021).

[5] X. Wang, H. Yuan, and X. Wang, Commun. Phys. **1**, 31 (2018).

[6] K. Wang, L. Qian, W. Chen, S.-C. Ying, G. Xiao, and X. Wu, Phys. Rev. B **99**, 184410 (2019).

[7] K. Wang, Y. Zhang, V. Bheemarasetty, S. Zhou, S.-C. Ying, and G. Xiao, Nat. Commun. **13**, 722 (2022).

[8] N. Nagaosa, J. Sinova, S. Onoda, A. H. MacDonald, and N. P. Ong, Rev. Mod. Phys. **82**, 1539 (2010).

[9] B. Dieny and M. Chshiev, Rev. Mod. Phys. **89**, 025008 (2017).

[10] G. Yu *et al.*, Nat. Nanotechnol. **9**, 548 (2014).

[11] B. Rodmacq, A. Manchon, C. Ducruet, S. Auffret, and B. Dieny, Phys. Rev. B **79**, 024423 (2009).

[12] M. Xu *et al.*, Phys. Rev. Lett. **124**, 187701 (2020).

[13] Y. Fan, K. Smith, G. Lüpke, A. Hanbicki, R. Goswami, C. Li, H. Zhao, and B. Jonker, Nat. Nanotechnol. **8**, 438 (2013).

[14] I. Dzyaloshinsky, J. Phys. Chem. Solids **4**, 241 (1958).

[15] T. Moriya, Phys. Rev. **120**, 91 (1960).

[16] L. Qian, K. Wang, Y. Zheng, and G. Xiao, Phys. Rev. B **102**, 094438 (2020).

[17] R. L. Conte *et al.*, Phys. Rev. B **91**, 014433 (2015).

[18] M. Cecot *et al.*, Sci. Rep. **7**, 968 (2017).

[19] J. J. Senkevich, T. Karabacak, D.-L. Bae, and T. S. Cale, J. Vac. Sci. Technol. **24**, 534 (2006).

[20] H. Zhang, S. Yamamoto, Y. Fukaya, M. Maekawa, H. Li, A. Kawasuso, T. Seki, E. Saitoh, and K. Takanashi, Sci. Rep. **4**, 4844 (2014).

[21] J. J. Colin, G. Abadias, A. Michel, and C. Jaouen, Acta Mater. **126**, 481 (2017).

[22] E. A. Ellis, M. Chmielus, and S. P. Baker, Acta Mater. **150**, 317 (2018).

[23] S. Sato, T. Inoue, and H. Sasaki, Thin Solid Films **86**, 21 (1981).

[24] Q. Hao, W. Chen, and G. Xiao, Appl. Phys. Lett. **106**, 182403 (2015).

[25] A. Patterson, Phys. Rev. **56**, 978 (1939).

[26] E. Zubler, J. Electrochem. Soc. **110**, 1072 (1963).

[27]    Y. A. Vodakov, A. Roenkov, M. Ramm, E. Mokhov, and Y. N. Makarov, Phys. Status Solidi B **202**, 177 (1997).

[28]    T. F. K. Lilov, S. O. S. Ohshima, and S. N. S. Nishino, Jpn. J. Appl. Phys. **40**, 6737 (2001).

[29]    E. Krikorian and R. Sneed, J. Appl. Phys. **37**, 3674 (1966).

[30]    L. Feinstein and R. Huttemann, Thin Solid Films **16**, 129 (1973).

[31]    R. Knepper, B. Stevens, and S. P. Baker, J. Appl. Phys. **100**, 123508 (2006).

[32]    T. Meier, M. Kronseder, and C. Back, Phys. Rev. B **96**, 144408 (2017).

[33]    I. Lemesh, F. Büttner, and G. S. Beach, Phys. Rev. B **95**, 174423 (2017).

[34]    S. Lemerle, J. Ferré, C. Chappert, V. Mathet, T. Giamarchi, and P. Le Doussal, Phys. Rev. Lett. **80**, 849 (1998).

[35]    A. B. Kolton, A. Rosso, T. Giamarchi, and W. Krauth, Phys. Rev. B **79**, 184207 (2009).

[36]    V. Jeudy, A. Mougin, S. Bustingorry, W. S. Torres, J. Gorchon, A. B. Kolton, A. Lemaître, and J.-P. Jamet, Phys. Rev. Lett. **117**, 057201 (2016).

[37]    K. Shahbazi, J.-V. Kim, H. T. Nembach, J. M. Shaw, A. Bischof, M. D. Rossell, V. Jeudy, T. A. Moore, and C. H. Marrows, Phys. Rev. B **99**, 094409 (2019).

[38]    W. S. Torres, R. D. Pardo, S. Bustingorry, A. B. Kolton, A. Lemaitre, and V. Jeudy, Phys. Rev. B **99**, 201201 (2019).

[39]    S.-Z. Lin, C. Reichhardt, C. D. Batista, and A. Saxena, Phys. Rev. B **87**, 214419 (2013).

[40]    O. I. Utesov, Phys. Rev. B **103**, 064414 (2021).

[41]    C. Song *et al.*, Adv. Funct. Mater. **31**, 2010739 (2021).

[42]    K. Wang, Y. Zhang, V. Bheemarasetty, S. Zhou, S.-C. Ying, and G. Xiao, Nat. Commun. **13**, 722 (2022).

[43]    A. Vansteenkiste, J. Leliaert, M. Dvornik, M. Helsen, F. Garcia-Sanchez, and B. Van Waeyenberge, AIP Advances **4** (2014).

[44]    A. Gurevich, Phys. Rev. B **42**, 4857 (1990).

[45]    S. Demirdiş, C. J. van der Beek, Y. Fasano, N. R. Cejas Bolecek, H. Pastoriza, D. Colson, and F. Rullier-Albenque, Phys. Rev. B **84**, 094517 (2011).

[46]    V. Sandu, A. M. Ionescu, G. Aldica, M. A. Grigoroscuta, M. Burdusel, and P. Badica, Sci. Rep. **11**, 5951 (2021).

[47]    G. D'anna, P. Gammel, H. Safar, G. Alers, D. Bishop, J. Giapintzakis, and D. Ginsberg, Phys. Rev. Lett. **75**, 3521 (1995).

[48]    C. Olson, C. Reichhardt, and F. Nori, Phys. Rev. Lett. **81**, 3757 (1998).

[49]    A. Marley, M. Higgins, and S. Bhattacharya, Phys. Rev. Lett. **74**, 3029 (1995).

[50]    S. Okuma, J. Inoue, and N. Kokubo, Phys. Rev. B **76**, 172503 (2007).

[51] I. Bloom, A. Marley, and M. Weissman, Phys. Rev. Lett. **71**, 4385 (1993).

[52] G. Grüner, A. Zawadowski, and P. Chaikin, Phys. Rev. Lett. **46**, 511 (1981).

[53] M. Yoshida, T. Sato, F. Kagawa, and Y. Iwasa, Phys. Rev. B **100**, 155125 (2019).